\documentclass[%
reprint,
 amsmath,amssymb,
 aps
pre,
floatfix,
]{revtex4-2}
\usepackage{amsthm}
\usepackage[version=4]{mhchem}
\usepackage{tikz-network}
\usetikzlibrary{decorations.markings} 

\theoremstyle{plain}
\newtheorem{remark}{Remark}
\newtheorem{prop}{Proposition}
\usetikzlibrary{math}
\usetikzlibrary[arrows]
\usepgflibrary{decorations.pathreplacing}

\newcommand{\tot}{\textrm{tot}}
\newcommand{\qss}{\textrm{qss}}

\newcommand{\bR}{ {\mathbb{R}} }

\newcommand{\Talpha}{ \tilde{\alpha}}
\newcommand{\Ttheta}{ \tilde{\theta}}
\newcommand{\TKd}{ \tilde{K}_d}

\newcommand{\kmeth}{ \frac{\Talpha_2}{\Ttheta_2} }
\newcommand{\gso}{ g_{\textrm{eq}}^{\textrm{MFL}} }

\newcommand{\eq}{ \textrm{eq} }
\newcommand{\Bso}{ B_{\textrm{eq}}^{\textrm{MFL}} }
\newcommand{\gse}{ g_{\textrm{eq}}^{\textrm{IT-MFL}} }
\newcommand{\Bse}{ B_{\textrm{eq}}^{\textrm{IT-MFL}} }
\renewcommand{\dim}{\textrm{dim}}
\newcommand{\Aq}{\tilde{A}_{\qss}(\tilde{B}_{\tot})}
\newcommand{\Tbt}{T(\tilde{B}_\tot)}
\newcommand{\Bt}{\tilde{B}_\tot}
\usepackage{graphicx}
\usepackage{dcolumn}
\usepackage{bm}

\begin{document}

\preprint{APS/124-QED}

\title{Promoter methylation in a mixed feedback loop circadian clock model}

\author{Turner Silverthorne$^{1,2}$, Edward Saehong Oh$^{2}$, and Adam R Stinchcombe$^{1}$}
\affiliation{
    $^1$Department of Mathematics, University of Toronto, Toronto, ON, Canada.\\
    $^2$The Krembil Family Epigenetics Laboratory, The Campbell Family Mental Health Research Institute, Centre for Addiction and Mental Health, Toronto, ON, Canada.}
\date{\today}

\begin{abstract}
We introduce and analyze an extension of the mixed feedback loop model of Fran\c{c}ois and Hakim. Our extension includes an additional promoter state and allows for reversible protein sequestration, which was absent from the original studies of the mixed feedback loop model. Motivated by experimental observations that link DNA methylation with circadian gene expression, we use our extended model to investigate the role of DNA methylation in the mammalian circadian clock. We extend the perturbation analysis of Fran\c{c}ois and Hakim to determine how methylation affects the presence and the periodicity of oscillations. We derive a modified Goodwin oscillator model as an approximation to show that although methylation contributes to period control, excessive methylation can abolish rhythmicity. 
\end{abstract}

\maketitle

\section{Introduction}
Epigenetic DNA modifications are far more dynamic than their traditional depiction \cite{OhPetronis2021,parry2020active}. Indeed, the modification status of cytosines (5-mC, 5-hmC, 5-fC and 5-caC) can vary significantly with various timescales: years/age \cite{HorvathSteve2018Dmba}, hours \cite{GabrielOh2018Cmec, GabrielOh2019Cooc, rulands2018genome}, and minutes \cite{Kangaspeska2008, Metivier2008}. Ongoing experimental efforts have focused on the relevance of epigenetic oscillations to biological function and phenotypes. In this paper, we focus on a particular instance of this general phenomenon: the influence of DNA methylation on the circadian clock. 

Both steady-state and oscillatory differences in the methylation status of clock genes in mammals have been detected in recent years. Azzi et al. entrained mice to a 22 hour day and found that after removing the entrainment cue, the mice retained a shortened circadian period and had differentially methylated clock genes \cite{azzi2014circadian}. From this observation, they hypothesized that DNA methylation of clock genes contributes to the plasticity of the mammalian circadian clock. More recently, circadian epigenomic studies have found evidence for 24-hour diurnal oscillations in cytosine modifications in human neutrophils \cite{GabrielOh2019Cooc}, as well as mouse liver and lung \cite{GabrielOh2018Cmec}. Interestingly, a portion of stochastic intra- and inter- individual epigenetic variation was accounted for by oscillations \cite{GabrielOh2019Cooc}, and sequences surrounding oscillating cytosine modifications were enriched in both canonical (CANNTG) and non-canonical (CANNNTG, GANNTG) enhancer elements \cite{GabrielOh2018Cmec} -- known as E-boxes -- which play key roles in regulation of circadian transcripts \cite{KoAndTakahashi2006}. These features of oscillating cytosine modifications suggest that oscillations in cytosine modifications are intricately linked to circadian transcriptomics, possibly by regulating the epigenetic status of E-box motifs.

Our aim in this paper is to add a mathematical perspective to these intriguing experimental findings. To this end, we focus on the PER transcription-translation feedback loop (TTFL), which is known to be of primary importance for rhythm generation in the mammalian circadian clock. We model this system using an extension of the mixed feedback loop (MFL) model of Fran\c{c}ois and Hakim~\cite{franccois2004design}.

The MFL model originated in \emph{in silico} studies of evolution \cite{franccois2004design}, and has since been found to be present in a variety of biologically important networks. For instance the circadian clocks of neuropsora \cite{loros2001genetic} and drosophila \cite{franccois2005core}, the p53-Mdm2 module \cite{bar2000generation}, and the \emph{E. coli} lactose operon \cite{monod1961general}. More recently, the MFL model has been used in several studies as a minimal model of the circadian clock \cite{kim2012mechanism,kim_molecular_2014,kim_protein_2016,wang_entrainment_2018}. Building on work of Kim and Forger \cite{kim2012mechanism}, who used the MFL model in their analysis of a detailed mammalian clock model, we add an additional promoter state to represent DNA methylation. Since the transcription rate of the new promoter state will be assumed to lie between the active and inactive transcription rates, we refer to our extension of the MFL model as the \emph{intermediate transcription rate} MFL model (IT-MFL model). In some parts of our analysis, we also consider reversible protein binding which was crucial to the work of Kim and Forger but was absent from the original papers on the MFL model.

Our analysis is divided into two parts. Inspired by the work of Fran\c{c}ois and Hakim, we begin by studying the system perturbatively. We extend their boundary layer analysis to derive approximate expressions for the period and bounds on the influence of methylation on the period in a limiting case of the model. A recurring theme in our perturbative analysis is that the MFL and IT-MFL models display qualitatively similar behaviour at dominant order provided that the transcription rate of the new promoter state is in fact \emph{intermediate}. In the second part, we extend the work of Kim and Forger~\cite{kim2012mechanism}. In addition to simulating a detailed model of the clock, they also used a modified form of the Goodwin oscillator to test their hypothesis that a balanced stoichiometry between activators and repressors was necessary for autonomous oscillations in the clock. We show how their model -- which takes the form of a monotone cyclic feedback (MCF) system -- can be obtained as an approximation of the IT-MFL model. We relax the assumption of constant promoter states to a quasi-steady state approximation and analyze the transcription function in this more general setting. Working in this approximation, our analysis reveals that although the period and its derivatives in parameter space are sensitive to methylation, excessive methylation can abolish rhythmicity. Numerical bifurcation analysis reveals that this loss of rhythmicity occurs through a supercritical Hopf bifurcation. The fact that the qualitative behaviour of our model is sensitive to slight differences in transcriptional regulation aligns with the general principle that even slight changes in transcriptional regulation can dramatically alter the behaviour of a genetic oscillator~\cite{dewoskin2014not}. 

\section{Mathematical model}\label{sec:mathmodel}
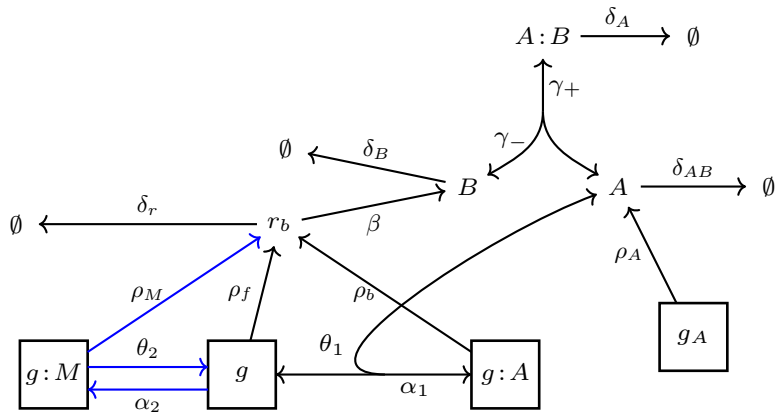
\begin{figure*}
    \centering
\begin{tikzpicture}[<->]

\Vertex[x=1.5,   y=0.5,   label=$g\!:\!M$  ,shape=rectangle,size=.9,color=none,fontscale=1.3]{gM}
\Vertex[x=1.5,   y=0.6,Pseudo]{gMupper}  
\Vertex[x=1.5,   y=0.3,Pseudo]{gMlower}  
\Vertex[x=4,     y=0.5,   label=$g$        ,shape=rectangle,size=.9,color=none,fontscale=1.3]{g}
\Vertex[x=5.2,   y=0.90,   label=$\theta_1$,Pseudo,fontscale=1.3]{theta1label}
\Vertex[x=8.3,     y=4.3,   label=$\gamma_+$,Pseudo,fontscale=1.3]{gammapluslabel}
\Vertex[x=7.58,     y=3.6,   label=$\gamma_-$,Pseudo,fontscale=1.3]{gammaminuslabel}
\Vertex[x=4.58,     y=3.5,   label=$\emptyset$,Pseudo,fontscale=1.3]{Bdecay}
\Vertex[x=6.3,   y=0.3,   label=$\alpha_1$,Pseudo,fontscale=1.3]{alpha1label}
\Vertex[x=4,     y=0.6,Pseudo]{gupper}  
\Vertex[x=4,     y=0.3,Pseudo]{glower}  
\Vertex[x=4.5,   y=2.5,   label=$r_b$        ,Pseudo,fontscale=1.3]{r}
\Vertex[x=7.5,   y=0.5,   label=$g\!:\!A$  ,shape=rectangle,size=.9,color=none,fontscale=1.3]{gsA}
\Vertex[x=10,    y=1, label=$g_A$      ,shape=rectangle,size=.9,color=none,fontscale=1.3]{gA}
\Vertex[x=6.5,   y=0.6,Pseudo]{gAupper}  
\Vertex[x=6.5,   y=0.3,Pseudo]{gAlower}  
\Vertex[x=7,     y=3,   label=$B$        ,Pseudo,fontscale=1.3]{B}
\Vertex[x=1,     y=2.5,   label=$\emptyset$,Pseudo,fontscale=1.3]{C2} 
\Vertex[x=11,    y=3,   label=$\emptyset$,Pseudo,fontscale=1.3]{D1} 
\Vertex[x=9,     y=3,   label=$A$        ,Pseudo,fontscale=1.3]{A}
\Vertex[x=8,     y=3,          ,Pseudo,fontscale=1.3]{D3}
\Vertex[x=8,     y=5,   label=$A\!:\!B$  ,Pseudo,fontscale=1.3]{AB}
\Vertex[x=10,    y=5,   label=$\emptyset$,Pseudo,fontscale=1.3]{E2}

\draw [->,thick,blue] (gM)  --node [midway,left,black]{$\rho_M$} (r);
\draw [->,thick]      (g)   --node [midway,left,black]{$\rho_f$} (r);
\draw [->,thick]      (gsA) --node [midway,left,black]{$\rho_b$} (r);
\draw [->,thick]      (r)   --node [midway,above,black]{$\delta_r$} (C2);
\draw [->,thick]      (B)   --node [midway,above,black]{$\delta_B$} (Bdecay);
\draw [->,thick]      (r)   --node [midway,below,black]{$\beta$} (B);
\draw [->,thick]      (gA)  --node [midway,left,black]{$\rho_A$} (A);
\draw [->,thick]      (8,4.0)  to [out=-90,in=30]  (B);
\draw [->,thick]      (8,4.0)  to [out=-90,in=150]  (A);
\draw [->,thick]      (8,4.0)  to [out=90,in=-90]  (AB);
\draw [->,thick, shorten >=0.01cm,shorten <=0.2cm]      (AB)  --node [midway,above,black]{$\delta_A$} (E2);
\draw [->,thick]      (A)   --node [midway,above,black]{$\delta_{AB}$} (D1);

\draw[->,thick,blue,shorten >=0.15cm,shorten <=0.15cm](gMupper) --node [midway,above,black]{$\theta_2$}  (gupper);
\draw[->,thick,blue,shorten >=0.15cm,shorten <=0.15cm](glower) --node  [midway,below,black]{$\alpha_2$} (gMlower);
\draw[<->,thick,black](g) -- (gsA);
\draw[->,thick,black](5.9,0.5) to [out=179,in=200]  (A);

\end{tikzpicture}
    \caption{(Colour online). Reaction diagram for the IT-MFL model. The activator $A$ is constitutively expressed, whereas the promoter of the target protein $B$ can be an in an active, inactive, or methylated state. Since $B$ binds to $A$, this sequestration mechanism forms a negative feedback loop. Reactions present in the original MFL model are drawn in black, and the reactions new to the IT-MFL model are drawn in blue.}
    \label{fig:fh_interaction}
\end{figure*}

An illustration of the MFL model is given in Fig.~\ref{fig:fh_interaction}, along with the extensions of the IT-MFL model. Tables~\ref{tab:parameters}-\ref{tab:paramtables} list the IT-MFL model's parameters and dynamic variables, respectively. The MFL model consists of an activator protein $A$ which interacts with a target protein $B$. The activator is assumed to be constitutively expressed whereas $B$ is regulated by a promoter that is active when bound to $A$ and otherwise inactive. Finally, $B$ can bind to $A$, sequestering it away from its promoters and thereby repressing its own transcription  \cite{buchler2008molecular,buchler2009protein}. The MFL model has been applied in several biological contexts, including the work of Kim and Forger on circadian rhythms~\cite{kim2012mechanism}. When studied in the context of the mammalian circadian clock, the activator $A$ and target protein $B$ represent CLOCK-BMAL1 and PER, respectively. Although the clock is made up of several interlocking feedback loops, the PER TTFL is a well-established starting point for a minimal model for analyzing the dynamics of this complex system~\cite{kim2012mechanism}.

We extend the MFL model in two ways: we add a new promoter state to represent methylation of the PER E-boxes and allow for the unbinding of $A$ and $B$. Some effects of the unbinding were discussed by Kim and Forger in a reduced form of the MFL model, but to the best of our knowledge this has not yet been studied in the full MFL model. As mentioned in the introduction, we assume the transcription rate corresponding to the methylated promoter state lies between the transcription rates of the active and inactive states. Although DNA methylation and demethylation are catalyzed by a variety of enzymes, the IT-MFL model assumes these reactions are operating with first-order kinetics. Explicit incorporation of the methylation and demethylation enzymes would be a natural next step to this work. Our analysis in this section follows the methodology from the original paper of Fran\c{c}ois and Hakim and so we adopt their notation for our model. 

\begin{table}
		\centering
            \begin{tabular}{||c|l||}\hline
                 Parameter & Meaning  \\ \hline
                    $\alpha_1   $& $E\to E:A$  reaction rate  \\ 
                    $\theta_1   $& $E:A \to E$ reaction rate  \\   
                    $\alpha_2   $& $E\to E:M$  reaction rate  \\ 
                    $\theta_2   $& $E:M \to E$ reaction rate  \\   
                    $\rho_M     $& \emph{per} transcr. rate (methylated E-boxes) \\   
                    $\rho_f     $& \emph{per} transcr. rate (inactive E-boxes) \\   
                    $\rho_b     $& \emph{per} transcr. rate (active E-boxes)  \\   
                    $\beta      $& rate of PER translation  \\ 
                    $\rho_A     $& activator expression rate \\  
                    $\gamma_+   $& $A+B \to A:B$  reaction rate \\ 
                    $\gamma_-   $& $A:B \to A+B$  reaction rate \\ 
                    $\delta_A   $& decay rate of activator \\  
                    $\delta_B   $& decay rate of B \\  
                    $\delta_{AB}$& decay rate of $A:B$  \\  
                    $\delta_r   $& decay rate of $r_b$ \\\hline 
            \end{tabular}
    \caption{Parameters in the dimensionful IT-MFL model. When chemical species $A$ and $B$ are bound to one another, we denote this by $A:B$.}
    \label{tab:parameters}
\end{table}
\begin{table}
		\centering
            \begin{tabular}{||c|l||}\hline
                    Dynamic variable & Meaning  \\ \hline
                    $[g] $   & inactive E-box conc.  \\ 
                    $[g:A]$  & active E-box conc.  \\ 
                    $[g:M]$  & methylated E-box conc.  \\   
                    $[r_b]$  & \emph{per} mRNA conc.  \\   
                    $[B] $   & PER protein conc. \\   
                    $[A] $   & activator protein conc. \\   
                    $[A:B] $ & per:activator conc. \\ \hline 
            \end{tabular}
    \caption{Dynamic variables in the dimensionful IT-MFL model.}
    \label{tab:paramtables}
\end{table}

\subsection{Parameters, dynamic variables, and governing equations}
Our notation is summarized in Table \ref{tab:paramtables}.  We use mass-action kinetics \cite{ingalls} to obtain the following governing equations for the promoter states
\begin{align}\label{eqn:firstofDim}
    \frac{d[g]}{dt}   &= \theta_1 [g:A] + \theta_2[g:M] - \alpha_1 [g] [A]  - \alpha_2 [g],\\
    \frac{d[g:M]}{dt} &= \alpha_2 [g] - \theta_2 [g:M],\\
    \frac{d[g:A]}{dt} &= \alpha_1 [g][A] - \theta_1 [g:A],\\
    g_{\tot} &= [g] + [g:M] + [g:A],
\end{align}
and similarly for the mRNA and protein concentrations
\begin{align}
    \frac{d[r_b]}{dt} &= \rho_f [g] + \rho_b [g:A] + \rho_M [g:M] - \delta_r [r_b],\\
    \frac{d[B]}{dt} &= \beta[r_b] - \delta_B [B] - \gamma_{+}[A][B]+ \gamma_{-}[A:B],\\
    \frac{d[A]}{dt} &= \rho_A - \gamma_{+}[A][B]  -\delta_A [A] +\theta_1 [g:A] \notag \\ 
    &\quad-\alpha_1[g][A] + \gamma_{-}[A:B],\\
    \frac{d[A:B]}{dt} &= \gamma_+ [A] [B] - \gamma_{-} [A:B] - \delta_{AB} [A:B]. \label{eqn:lastofDim}
 \end{align}
For the remainder of this section we use the same non-dimensionalization procedure as Fran\c{c}ois and Hakim to derive a dimensionless form of equations Eqs.~\eqref{eqn:firstofDim}-\eqref{eqn:lastofDim}. Let $\tilde{t}$ be the dimensionless time $\tilde{t} := \delta_r t$ and write $\dot{u}: = \frac{du}{d\tilde{t}}$ for $u \in C^{1}(\bR)$, a function of dimensionless time. We normalize the promoter states so that $g_{\tot} =1$ and obtain dimensionless equations for their time evolution
 \begin{align}
     \dot{g} &= \tilde{\theta}_1\left( \left(1-g-g_M \right) + \frac{\tilde{\theta}_2}{\tilde{\theta}_1}g_M - g \frac{A}{A_0}\right) - \tilde{\alpha}_2 g \label{eqn:modelstart}, \\
     \dot{g}_M &= \tilde{\alpha}_2 g - \tilde{\theta}_2 g_M,
 \end{align}
in which $g = [g]/g_{\tot}$ and $g_M = [g:M]/g_{\tot}$. $g_A = [g_A]/g_{\tot}$ is given by $g_A =1-g-g_M$. We rescale the protein concentrations so that $A = \sqrt{\gamma_{+}/\rho_A} [A],B = \sqrt{\gamma_{+}/\rho_A} [B], A_B = \sqrt{\gamma_{+}/\rho_A} [A:B], r = \sqrt{\gamma_{+}/\rho_A} [r_b]$ to obtain 
 \begin{align}
     \dot{r} &= \rho_0 g + \rho_1 (1-g-g_M) + \rho_2 g_M - r ,\\ 
     \dot{A}     &= \frac{1}{\delta}\left( 1-A \cdot B\right)  - d_a A   + \TKd A_B\notag \\
     &\quad+ \mu\tilde{\theta}_1 \left( (1-g-g_M)- g \frac{A}{A_0}\right), \\
     \dot{B}     &=  \frac{1}{\delta} \left(r- A\cdot B\right) - d_b B + \TKd A_B,\\
     \dot{A}_B   &= \frac{A\cdot B}{\delta} -  d_{AB}A_B
     \label{eqn:modelend}.
 \end{align}

Altogether, the IT-MFL model consists of the parameters in Table~\ref{fig:dimlessModelParams} and the six dynamic variables $g,g_M,r,A,B, A_{B}$ that evolve according to Eqs.~\eqref{eqn:modelstart}~-~\eqref{eqn:modelend}. It should be noted that aside from the nonlinearity introduced by the sequestration of $A$ by $B$, the dynamics of the MFL and IT-MFL models are linear. This weak nonlinearity has made the MFL model attractive for stochastic extensions \cite{karapetyan2015role}.
\begin{table}[htp]
    \centering
    \begin{tabular}{||c|c|l||}\hline
         Parameter & Formula & Meaning  \\ \hline
$\rho_0               $ & $ \frac{\beta \rho_f}{\rho_A \delta_r} $     & $B$ transcr. rate (inactive E-boxes)\\
$\rho_1               $ & $ \frac{\beta \rho_b}{\rho_A \delta_r} $     & $B$ transcr. rate (active E-boxes)\\
$\tilde{\theta}_1     $ & $ \frac{\theta_1}{\delta_r}            $     & $g:A \to g$ rate / \emph{per} mRNA decay \\
$\delta               $ & $  \frac{\delta_r}{\sqrt{\rho_A\gamma_{+}}}$ & scaled $B$ mRNA decay \\
$d_a                  $ & $ \frac{\delta_A}{\delta_r}$                 & normalized activator decay \\
$d_b                  $ & $ \frac{\delta_B}{\delta_r}$                 & normalized $B$ decay \\
$d_{AB}                  $ & $ \frac{\delta_{AB}}{\delta_r}$           & normalized  $A:B$ decay \\
$\mu                  $ & $ \sqrt{\frac{\gamma_{+}}{\rho_A}}$          & activator consumption / production \\
$A_0                  $ & $ \frac{\theta}{\alpha}\sqrt{\frac{\gamma_{+}}{\rho_A}}$ & dimensionless critical binding scale \\ 
$\TKd                 $ & $ \frac{K_d \gamma_+}{\delta_r}$ & dissociation constant for $A,B$ binding \\ \hline 

$\tilde{\alpha}_2     $ & $ \frac{\alpha_2}{\delta_r} $                & scaled rate of promoter demethylation\\
$\tilde{\theta}_2     $ & $ \frac{\theta_2}{\delta_r} $                & scaled rate of promoter methylation\\
$\rho_2               $ & $ \frac{\beta \rho_M}{\rho_A \delta_r} $     & \emph{per} transcr. rate (methylated E-boxes) \\ \hline
    \end{tabular}
    \caption{Parameters for the dimensionless form of the MFL model. The final three parameters are new to the IT-MFL model.}
    \label{fig:dimlessModelParams}
\end{table}
Of the several dimensionless parameters listed in Table~\ref{fig:dimlessModelParams}, we will be most interested in the transcription rates $\rho_0,\rho_1,\rho_2$, the methylation and demethylation rates $\Talpha_2,\Ttheta_2$, and the timescale ratio $\delta$. 
\section{Analysis}
Our aim is to determine how the intermediate promoter state affects the stability and period of the IT-MFL model. Building on work of Fran\c{c}ois and Hakim, we begin by expanding the system perturbatively in $\delta$. This reveals that in the small $\delta$ regime, the behaviour of the MFL and IT-MFL models are qualitatively similar provided that
\begin{equation}\rho_0<\rho_2<\rho_1.\end{equation}
In the latter part of this section, we focus on the circadian setting and explicitly show how the Kim-Forger model can be derived from the IT-MFL model. We also relax an assumption commonly employed in this derivation, and explore the consequences of this choice. In its reduced form, the IT-MFL model is a monotone cyclic feedback system and therefore obeys a  generalization of the Poincar\'{e}-Bendixson theorem in this regime. This structure plays a key role in our bifurcation analysis of the reduced model. Using parametric sensitivity analysis, we show that although the period is not particularly sensitive to the methylation parameters, they do play a nontrivial role in determining the sensitivity of the period to the other parameters in the model. In general, the period becomes most sensitive as the model approaches a Hopf bifurcation rendering its equilibrium stable.

\subsection{Equilibrium uniqueness conditions}\label{sec:monostab}
The assumption that $A$ is an activator $(\rho_0<\rho_1)$ eliminates the possibility for multistability in the MFL model. Fran\c{c}ois and Hakim showed this by reducing the equilibrium equations of the MFL model to
\begin{equation}
1 = \delta d_a A +  \frac{A(\rho_1 A + \rho_0 A_0)}{(A+A_0)(A+\delta d_b)}.\label{eqn:FHsingle}
\end{equation}
The right hand side of Eq.~\eqref{eqn:FHsingle} vanishes when $A=0$, tends to infinity as $A\to\infty$, and is monotonic in $A$ when $\rho_0<\rho_1$. It follows that there is a unique non-negative value of $A$ that satisfies Eq.~\eqref{eqn:FHsingle} and determines the steady state of the system. A similar property is true of the IT-MFL model. We find that at steady-state, Eqs.~\eqref{eqn:modelstart}-\eqref{eqn:modelend} reduce to
\begin{equation}\label{eqn:Asteady}
  1 = \delta d_a A + \frac{A\left(1-\frac{\TKd^2}{d_{AB}^2}\right)\left( \rho_1 A + \left( \rho_0 + \frac{\rho_2 \Talpha_2}{\Ttheta_2} \right)A_0\right)}{\left(A+A_0\left(1 + \kmeth \right)  \right)\left(A+\delta d_b \left( 1 + \frac{\TKd}{d_{AB}} \right)\right)} .
\end{equation}
Note that Eq.~\eqref{eqn:FHsingle} is recovered from Eq.~\eqref{eqn:Asteady} in the limit of no methylation $(\tfrac{\Talpha_2}{\Ttheta_2}\to 0)$ and tight activator-target binding $(\TKd \to 0)$. If we assume the additional promoter state has an intermediate transcription rate $(\rho_0<\rho_2<\rho_1)$ then the right hand side of Eq.~\eqref{eqn:Asteady} is monotonic in $A$ and there remains a unique non-negative solution to the system's equilibrium equation. 

When the conditions for a unique non-negative equilibrium are not satisfied, one can proceed algebraically or perturbatively. The algebraic approach taken by Nagy produces a closed-form parameterization of the boundary between unique and multi equilibria in the parameter space of the MFL model \cite{nagy2018,simon1999constructing}. Although this approach still applies to the IT-MFL model, the expressions are more cumbersome and do not give much intuition on differences in the stability boundary in the two models. On the other hand, the perturbative approach taken by Fran\c{c}ois and Hakim is informative when applied to the IT-MFL model. Observe that Eq.~\eqref{eqn:Asteady} can be expanded perturbatively in $\delta$. Using the method of dominant balance \cite{bender2013advanced}, we find that $A$ may take low, medium, or high steady state values  
\begin{align}
   A_1 &:= \delta \frac{d_b\left(1+\kmeth\right) \left( 1 + \frac{\TKd}{d_{AB}}\right)}{ \rho_0 -1 + \kmeth(\rho_2 -1 ) } + \mathcal{O}(\delta^2), \label{eqn:firstExtended}\\ 
   A_2 &:= \frac{ A_0 \left( \rho_0-1 + \frac{\Talpha_2}{\Ttheta_2}(\rho_2-1) \right)}{1-\rho_1} + \mathcal{O}(\delta),\\ 
    A_3 &:= \frac{1-\rho_1}{\delta d_a}+ \mathcal{O}(1)\label{eqn:lastExtended}.
\end{align}
If $\rho_0<\rho_2< \rho_1$ then exactly one of $A_1,A_2,A_3$ will be non-negative, and thus we have not contradicted the monostability condition. Also note that the equilibrium solutions in Eqs.~\eqref{eqn:firstExtended}-\eqref{eqn:lastExtended} reduce to those found by Fran\c{c}ois and Hakim in the no methylation and tight activator-target binding limit.

\subsection{Linear stability analysis}\label{sec:lsafullmodel}
We now focus on the case where  $A$ is an activator and the new promoter state is intermediate $(\rho_0<\rho_2<\rho_1)$. From the previous section, we know there is a unique non-negative equilibrium $A_{\eq}=A_2$ in this case. We use the subscript $\textrm{eq}$ to refer to the steady-state value of a dynamic variable corresponding to the equilibrium $A(t)=A_\eq$. When linearized at $A_{\eq}$, we see in Table~\ref{tab:A2eigenvals} that the dominant order terms in the eigenvalues of the MFL and IT-MFL models are most easily compared when expressed in terms of the $g$ and $B$ steady-states. The first row of Table~\ref{tab:A2eigenvals} summarizes the findings of Fran\c{c}ois and Hakim, and the second row contains our extension of their calculations. Both models possess eigenvalues proportional to each of the cubic roots of unity, along with an eigenvalue $\lambda_4$ which is stable for all parameter values.  

Two additional eigenvalues are present in the the IT-MFL model. Under the assumption that $\rho_0 < \rho_2 \leq \rho_1$, we can show the $\lambda_5^{\textrm{IT-MFL}}$ eigenvalue is stable at dominant order.
\begin{prop}\label{prop:only} If $\rho_0 < \rho_2 \leq \rho_1$ then $\lambda_5^{\textrm{IT-MFL}} <0$ to dominant order as $\delta \to 0$.
\end{prop}
\begin{table*}
    \centering
    \begin{tabular}{||c|c|c||}\hline
         & Steady-state & Eigenvalues  \\ \hline
         MFL
         & {$\!\begin{aligned} 
                \gso &=  \frac{A_0}{A_{\eq} + A_0} \\
                \Bso &= \frac{\rho_1 A_{\eq} + \rho_0 A_0}{(A_{\eq}+A_0)(A_{\eq}+\delta d_b)}. 
           \end{aligned}$}
         & {$\!\begin{aligned} 
                \lambda^{\textrm{MFL}}_i &= - \omega_i \left( \frac{\Ttheta_1 \gso A_{\eq}(\rho_1-\rho_0)}{\delta A_0(A_{\eq}+\Bso)} \right)^\frac{1}{3} + \mathcal{O}(1)\\
                \lambda^{\textrm{MFL}}_4 &=\frac{-(A_{\eq} + \Bso)}{\delta}+ \mathcal{O}(1) 
           \end{aligned}$}\\ \hline
               
         IT-MFL 
         & {$\!\begin{aligned} 
                \gse &=  \frac{A_0}{A_{\eq} + A_0\left(1+\kmeth\right)} \\
                \Bse &= \frac{\left( 1 + \frac{\TKd}{d_{AB}} \right)\left(\rho_1 A_{\eq} + \left( \rho_0 + \frac{\rho_2 \Talpha_2}{\Ttheta_2} \right)A_0\right)}{\left(A_{\eq}+A_0\left(1 + \kmeth \right)  \right)\left(A_{\eq}+\delta d_b\left( 1 + \frac{\TKd}{d_{AB}}\right)\right)}. 
           \end{aligned}$}
         & {$\!\begin{aligned} 
                \lambda^{\textrm{IT-MFL}}_i &= - \omega_i \left( \frac{\Ttheta_1 \gse A_{\eq}(\rho_1-\rho_0)}{\delta A_0(A_{\eq}+\Bse)} \right)^\frac{1}{3} + \mathcal{O}(1)\\
                \lambda^{\textrm{IT-MFL}}_4 &=\frac{-(A_{\eq} + \Bse)}{\delta}+ \mathcal{O}(1) \\
                \lambda^{\textrm{IT-MFL}}_5 &=\frac{ \left(1 + \kmeth \right)\rho_1-\left( \rho_0 + \kmeth\rho_2\right) }{\frac{1}{\Ttheta_2}\left( \rho_{0}-\rho_{1}\right) } + \mathcal{O}(\delta)\\
                \lambda^{\textrm{IT-MFL}}_6 &=-d_{AB} + \mathcal{O}(\delta)
           \end{aligned}$}\\\hline
    \end{tabular}
    \caption{Steady-states and eigenvalues corresponding to the linearization of the MFL and IT-MFL models at the $A_\eq=A_2$ steady-state. $\omega_i$ denotes the $i$-th cubic root of unity for $i=1,2,3$. The additional eigenvalues in the IT-MFL model will be negative provided that $\rho_0< \rho_2<\rho_1$.}
    \label{tab:A2eigenvals}
\end{table*}
A proof of Prop.~\ref{prop:only} is given in the supplemental material. The other new eigenvalue $\lambda_6^\textrm{IT-MFL}$ is always stable to dominant order since $d_{AB}\geq 0$. We see that the qualitative behaviour -- the linear stability and number of non-negative equilibria -- is similar between the MFL and IT-MFL models provided that $\rho_0<\rho_2<\rho_1$. A natural extension of this work would be to see if this phenomenon persists in $n$-promoter state models.
\subsection{Period estimation}\label{sec:limitperiod}
\begin{figure*}
    \centering
\begin{tikzpicture}
    \tikzmath{\x = 6;
    \l1 = -\x; 
    \l2 = -6*\x/7;
    \l3 = -1*\x/7;
    \l4 = 0;
    \l5 = \x*2/7;
    \l6 = \x*5/7;
    \l7 = \x;}
    \draw [very thick] (\l1,-1) -- (\l1,1);
    \draw [very thick,dashed] (\l2,-1) -- (\l2,1);
    \draw [very thick,dashed] (\l3,-1) -- (\l3,1);
    \draw [very thick] (\l4,-1) -- (\l4,1);
    \draw [very thick,dashed] (\l5,-1) -- (\l5,1);
    \draw [very thick,dashed] (\l6,-1) -- (\l6,1);
    \draw [very thick] (\l7,-1) -- (\l7,1);
    \node [align=left] at (\l1,1.25) {$t=0$};
    \node [align=left] at (\l1 - 0.5*\l1+0.5*\l2,0) {$\textrm{BL}_4$};
    \node [align=left] at (\l3 - 0.5*\l3+0.5*\l4,0) {$\textrm{BL}_1$};
    \node [align=left] at (\l4 - 0.5*\l4+0.5*\l5,0) {$\textrm{BL}_2$};
    \node [align=left] at (\l6 - 0.5*\l6+0.5*\l7,0) {$\textrm{BL}_3$};
    \node [align=left] at (\l4,1.25) {$t=t_1$};
    \node [align=left] at (\l7,1.25) {$t=t_2$};
    \draw[very thick, decorate, decoration={brace,mirror},yshift=-1em ]  (\l1*.97,-1) -- node[below=0.4ex,align=center]{Phase I \\ (High A, Low B)}  (\l4-.05*\x,-1); 
    \draw[very thick, decorate, decoration={brace,mirror},yshift=-1em ]  (\l4+.05*\x,-1) -- node[below=0.4ex,align=center] {Phase II \\ (High B, Low A)}  (\l7*.97,-1); 
    \draw[very thick, decorate, decoration={brace},yshift=1.8em ]  (\l1,1) -- node[above=0.4ex] {$\mathcal{O}(\sqrt{\delta})$}  (\l2,1); 
    \draw[very thick, decorate, decoration={brace},yshift=1.8em ]  (\l3,1) -- node[above=0.4ex] {$\mathcal{O}(\sqrt{\delta})$}  (\l4,1); 
    \draw[very thick, decorate, decoration={brace},yshift=1.8em ]  (\l4,1) -- node[above=0.4ex] {$\mathcal{O}(\delta)$}  (\l5,1); 
    \draw[very thick, decorate, decoration={brace},yshift=1.8em ]  (\l6,1) -- node[above=0.4ex] {$\mathcal{O}(\delta)$}  (\l7,1); 
\end{tikzpicture}
    \caption{Structure of the boundary layers in a limit-cycle solution to the MFL model. Boundary layers $\textrm{BL}_2$ and $\textrm{BL}_3$ form when the system transitions from a phase of high A to high B concentration. Boundary layers $\textrm{BL}_1$ and $\textrm{BL}_4$ appear as the quasi-steady state approximation for $g$ breaks down.}
    \label{fig:BLTstructure}
\end{figure*}
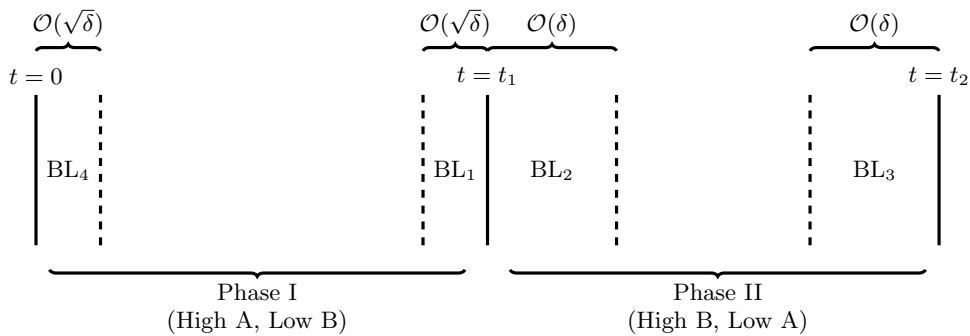
\begin{figure}
    \centering
    \includegraphics{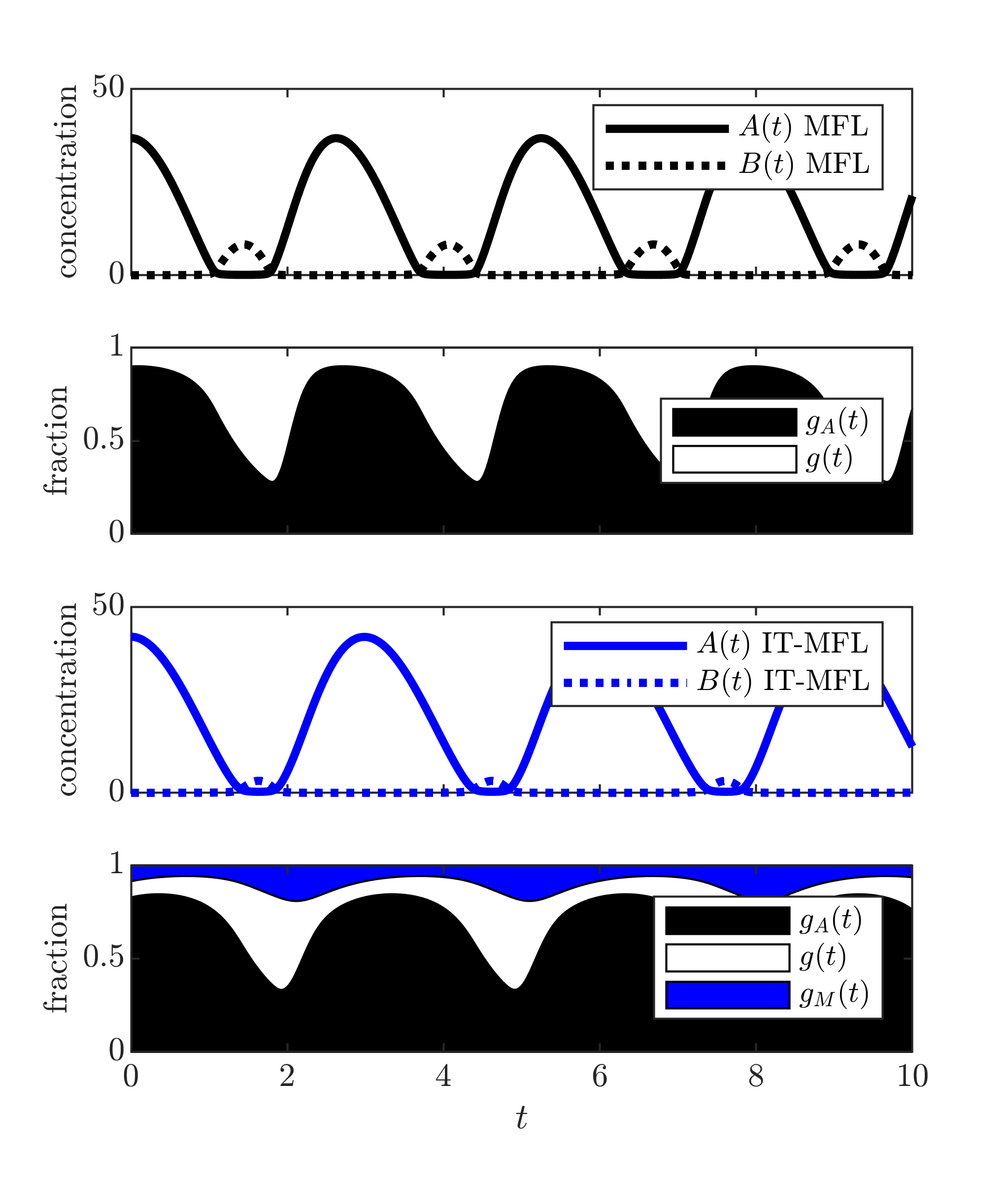}
    \caption{(Colour online). Oscillatory solutions to the MFL and IT-MFL models. In both cases, the oscillatory solutions decompose into phases of high-$A$/low-$B$ and high-$B$/low-$A$ concentration. Promoter states are shown for both models. Simulation parameters: $\delta=3\cdot 10^{-3},$ $\rho_0=0$, $\rho_1=1.45$, $\Ttheta_1=1.33$, $d_a=0.33$, $d_b=0.33$, $\mu =0.31$, $A_0 = 4$, $\rho_2 = 0$, $\Talpha_2=1$, $\Ttheta_2 =2$. }
    \label{fig:timestepABonly}
\end{figure} 
The numerical simulations in Fig.~\ref{fig:timestepABonly} show it is possible to change the period of the IT-MFL model by only altering the methylation parameters. To analyze such solutions, we make a change of variable $A\to \delta a$ for the high-$A$/low-$B$ phase of the limit cycle and $B\to \delta b$ for the high-$B$/low-$A$ phase. Following this substitution, we approximate the governing equations to lowest nontrivial order in $\delta$. In this approximation, some variables are left in steady-state and others obey linear differential equations. The steady state values and linear equations for each phase are summarized in the first two rows of Table~\ref{tab:BLTsummary}. Imposing continuity of the solution across the boundary layers, depicted in Fig.~\ref{fig:BLTstructure}, produces a system of nonlinear equations that determines the constants of integration and the durations of each phase. For both models, the system of nonlinear equations can be separated into boundary conditions and a closed system of equations implicitly determining the period. These systems are listed in the third and fourth rows of Table~\ref{tab:BLTsummary}.
\begin{table*}
    \centering
    \begin{tabular}{||c|c|c||}\hline
            & MFL & IT-MFL \\\hline
         Phase I & {$\!\begin{aligned} 
             g &= \frac{\delta A_0}{a}\\
             B &=  \frac{\delta r}{a}\\
             \dot{r} &=  \rho_1 - r\\
             \dot{a} &= 1 - r - d_a a.
           \end{aligned}$}
         & {$\!\begin{aligned} 
             g &= \frac{\delta A_0}{a} \left(1 - \left( 1 - \tfrac{\Ttheta_2}{\Ttheta_1}\right)g_{M} \right)\\
             B &= \frac{\delta r_0}{a} \\
             \dot{g}_M &= - \Ttheta_2 g_M\\
             \dot{r} &=  \rho_1(1-g-g_M ) + g_M\rho_2 - r\\
             \dot{a} &= 1 - r - d_a a.
           \end{aligned}$} \\ \hline 
         Phase II &{$\!\begin{aligned} 
             A &= \frac{\delta}{b}  \\
             \dot{g} &= \Ttheta_1(1-g)\\
             \dot{r} &= \rho_0 g + \rho_1(1 - g ) - r  \\
             \dot{b} &= r - 1 - d_b b.
           \end{aligned}$}
         & {$\!\begin{aligned} 
             A &= \frac{\delta}{b} \label{eqn:phase2ext_first} \\
             \dot{g} &= \Ttheta_1\left( 1 - g -g_M +\tfrac{\Ttheta_2}{\Ttheta_1}g_M \right) -\Talpha_2 g\\
             \dot{g}_M &=  \Talpha_2 g - \Ttheta_2 g_M \label{eqn:phase2ext_second} \\
             \dot{r} &= \rho_0 g + \rho_1(1 - g -g_M)+ g_M \rho_2 - r  \\
             \dot{b} &= r - 1 - b d_b\label{eqn:phase2ext_last}.
           \end{aligned}$}\\ \hline
         \vtop{\hbox{\strut Boundary conditions}\hbox{\strut $\mathcal{O}(\delta^0)$ estimate}}
         &  {$\!\begin{aligned}
            r_I(0) &= r_1, \;\;   a_I(0) = 0,\\
            g_{II}(0) &= 0, \;\;  r_{II}(0) = r_2,\\
            b_{II}(0) &= 0
         \end{aligned}$}
         &  {$\!\begin{aligned}
            r_I(0) &= r_1, \;\;   a_I(0) = 0, \;\; g_{M,I}(0) = g_{M1},\\
            g_{II}(0) &= 0, \;\;  g_{M,II}(0)= g_{M2}, \;\; r_{II}(0) = r_2,\\
            b_{II}(0) &= 0
         \end{aligned}$} \\\hline
         $\mathcal{O}(\delta^0)$ Estimate & {$\!\begin{aligned} 
             r_I(t_1) &= r_2,\;\; a_I(t_1) = 0, \\
             r_{II}(t_2) &= r_1,\;\; b_{II}(t_2) = 0 
           \end{aligned}$}  &  {$\!\begin{aligned} 
             r_I(t_1) &= r_2, \;\; a_I(t_1) = 0, \;\; g_{M,I}(t_1) = g_{M2},\\
             r_{II}(t_2) &= r_1, \;\; b_{II}(t_2) = 0, \;\; g_{M,II}(t_2) = g_{M1}
           \end{aligned}$}  \\ \hline 
         \vtop{\hbox{\strut Boundary conditions}\hbox{\strut $\mathcal{O}(\sqrt{\delta})$ estimate}}
         &  {$\!\begin{aligned}
            r_I(0) &= r_1, \;\;   a_I(0) = 0,\\
            g_{II}(0) &= \sqrt{\frac{\pi A_0 \Ttheta_1 \delta}{2(r_2 -1)}}, \;\;  r_{II}(0) = r_2,\\
            b_{II}(0) &= 0
         \end{aligned}$}
         &  {$\!\begin{aligned}
            r_I(0) &= r_1, \;\;   a_I(0) = 0, \;\; g_{M,I}(0) = g_{M1},\\
            g_{II}(0) &=\left( \Ttheta_1 + (\Ttheta_2 - \Ttheta_1)g_{M2}\right)\sqrt{\frac{\pi A_0 \delta}{2\Ttheta_1 (r_2-1)}} ,\\
           g_{M,II}(0)&= g_{M2}, \;\; r_{II}(0) = r_2,\\
            b_{II}(0) &= 0
         \end{aligned}$} \\\hline
         $\mathcal{O}(\sqrt{\delta})$ Estimate & {$\!\begin{aligned} 
             r_I(t_1) &= r_2,\;\; a_I(t_1) = 0, \\
             r_{II}(t_2) &= r_1 + g_1 (\rho_1-\rho_0)\sqrt{\frac{\pi}{4 \kappa_1}}\\
             b_{II}(t_2) &= 0 
           \end{aligned}$} 
         &  {$\!\begin{aligned}
             r_I(t_1) &= r_2, \;\; a_I(t_1) = 0, \;\; g_{M,I}(t_1) = g_{M2},\\
             r_{II}(t_2) &= r_1 + g_1 (\rho_1-\rho_0)\sqrt{\frac{\pi}{4 \kappa_1}}, \\
             b_{II}(t_2) &= 0, \;\; g_{M,II}(t_2) = g_{M1}
         \end{aligned}$} \\\hline
    \end{tabular}
    \caption{Summary of the nonlinear equations derived in the lowest order and dominant order period estimates. The rescaled variables are given by $a := \delta A$ and $b:=\delta B$ and the parameters $\kappa_1 := \frac{\Ttheta_1(1-r_1)}{2A_0\delta}$, and $g_1 := 1-e^{-\Ttheta_1 t_2}$ .}
    \label{tab:BLTsummary}
\end{table*}

\begin{figure}
    \centering
    \includegraphics{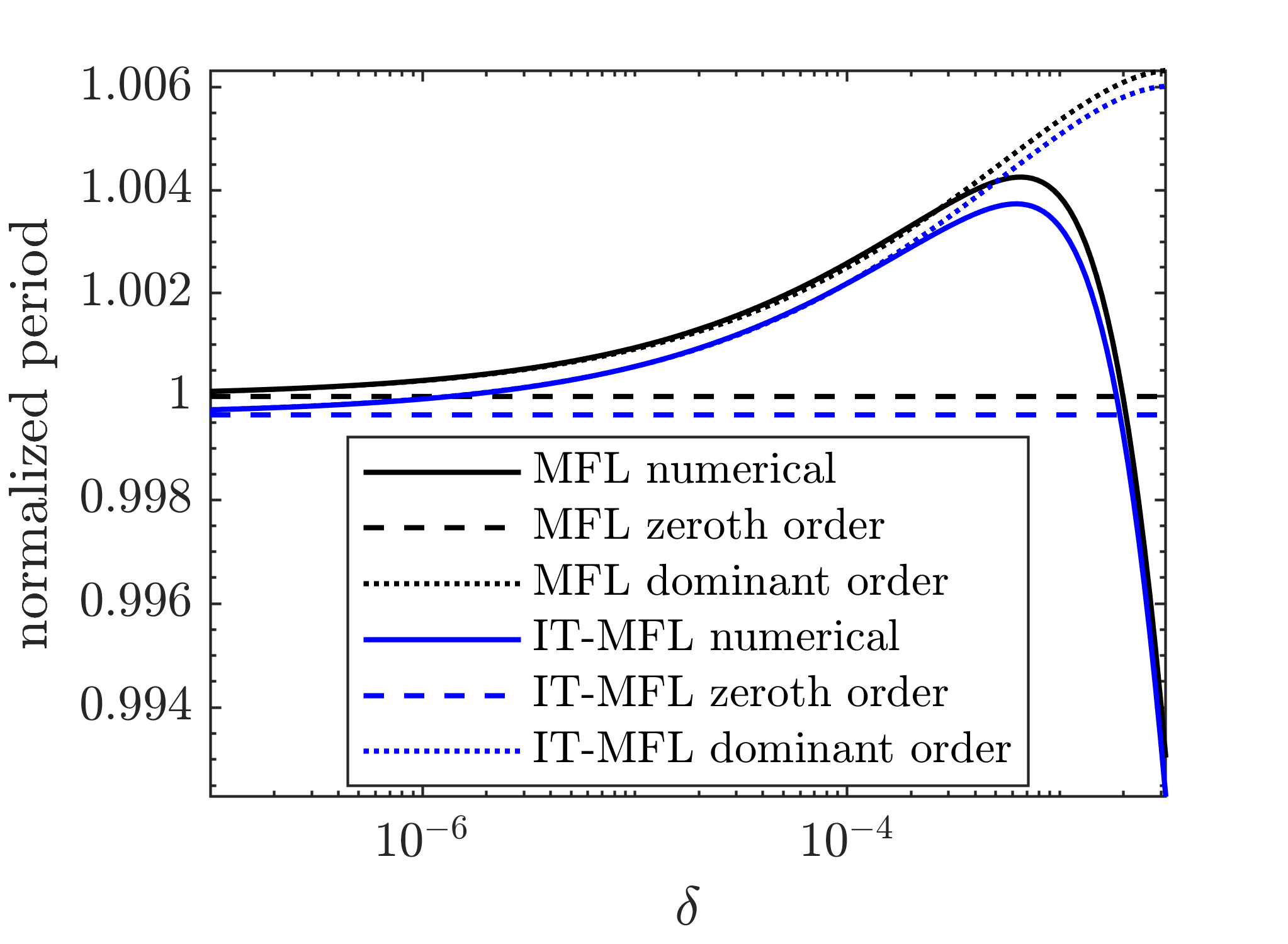}
    \caption{(Colour online). Comparison of numerically and asymptotically estimated periods. The period is expressed as a ratio with its limiting value as $\delta \to 0$. Parameters for the intermediate promoter state have been selected so its influence is weak $(\rho_2=0,\; \Talpha_2 \ll \Ttheta_2)$. Parameters: $\rho_0=0$, $\rho_1=10.45$, $\Ttheta_1=1.33$, $d_a=0.33$, $d_b=0.33$, $\mu =0.31$, $A_0 = 4$, $\rho_2 = 0$, $\Talpha_2=1.0$, $\Ttheta_2 =10$.}
    \label{fig:zerothasymptotic}
\end{figure}

When the influence of the smaller boundary layers are included, the $\mathcal{O}(\delta^0)$ period estimate is improved to an $\mathcal{O}(\sqrt{\delta})$ estimate. These corrections result in updated boundary conditions and an updated system of nonlinear equations, given in the final two rows of Table~\ref{tab:BLTsummary}. Comparing the $\mathcal{O}(\delta^0)$ and $\mathcal{O}(\sqrt{\delta})$ estimates, we see that the only change in boundary conditions is for the $g_{II}(t)$ solution and the only change in the period equations is for $r_{II}(t)$. Even in the case of the $\mathcal{O}(\delta^0)$ estimate in the MFL model, the system of implicit equations given in Table~\ref{tab:BLTsummary} cannot be solved exactly. Fig.~\ref{fig:zerothasymptotic} shows a strong agreement between a fully numerical simulation and a numerical solution of the nonlinear system of equations from Table~\ref{tab:BLTsummary}.

Continuing to follow the methodology of Fran\c{c}ois and Hakim, we make two additional simplifying assumptions to obtain approximate expressions for the $\mathcal{O}(\delta^0)$ period estimate. First, we assume the decay rate of $B$ is smaller than the decay rate of its mRNA. Second, we assume the transcription rate corresponding to the active promoter state is larger than the expression rate of $A$. When expressed in terms of the model parameters, these two assumptions give us  $d_b=0$ and $\rho_1 \gg 1$. When $d_b=0$, the expressions for the solutions in each phase of the limit cycle become much simpler. Under the assumption $\rho_1\gg 1$, it is justifiable to Taylor expand the system of implicit equations in powers of $\frac{1}{\rho_1}$. In addition, it can be verified by simulation and later confirmed by the approximate formulas in Table~\ref{tab:limEstimates} that $t_1$ decreases as $\rho_1$ increases. This allows us to also Taylor expand the implicit equations in powers of $t_1$. Table~\ref{tab:limEstimates} summarizes the result of Taylor expanding the implicit equations from the $\mathcal{O}(\delta^0)$ estimate from Table~\ref{tab:BLTsummary} in powers of $\frac{1}{\rho_1}$ and  $t_1$, and neglecting any terms that are exponentially small in $t_2$. Further details on this calculation are given in the supplemental material. Fig.~\ref{fig:limitingper} shows reasonable agreement between numerical period estimation and the approximate expressions from Table~\ref{tab:limEstimates}. Since we have Taylor expanded in powers of $\frac{1}{\rho_1}$, note that we expect to see agreement as $\rho_1$ gets large in Fig.~\ref{fig:limitingper}.  
\begin{table*}
    \centering
    \begin{tabular}{||l|l|l||}\hline
         Equation                 & MFL & IT-MFL  \\ \hline
        $a_I(t_1) = 0$            & $ t_1 = \frac{2(1-\rho_0)}{\rho_1} $ & $ t_1 =  \frac{2(1-\rho_0)}{\rho_1} + \frac{2 \left(1-\rho_2 \right)\frac{\Talpha_2}{\Ttheta_2}}{\rho_1}$ \\
        $r_I(t_1) = r_2$          & $ r_2 = 2 - \rho_0 $ & $r_2 =\frac{2-\rho_0 + (2-\rho_2)\frac{\Talpha_2}{\Ttheta_2} }{1 + \frac{\Talpha_2}{\Ttheta_2} }$ \\
        $r_{II}(t_2) = r_1$         & $ r_1 = \rho_0$ & $r_1 = \frac{\rho_0 + \rho_2\frac{\Talpha_2}{\Ttheta_2}}{1 +\frac{\Talpha_2}{\Ttheta_2}}$ \\
        $b_{II}(t_2) = 0$         & $ t_2 = 2 + \tfrac{\rho_1-\rho_0}{\Ttheta_1(1-\rho_0)}$ & $t_2 =  2 +  \frac{\rho_1-\rho_0 + \frac{\Talpha_2}{\Ttheta_2}\left(\rho_1- \rho_2 \right)}{\Ttheta_1\left( 1 - \rho_0 + (1-\rho_2)\frac{\Talpha_2}{\Ttheta_2}\right)\left(1 + \frac{\Talpha_2}{\Ttheta_2} \right) }$\\
        $g_{M,I}(t_1)= g_{M2}$   &  N/A & $g_{M2} = \frac{\Talpha_2}{\Talpha_2 + \Ttheta_2}$\\
        $g_{M,II}(t_2) = g_{M1}$ &  N/A & $g_{M1} = \frac{\Talpha_2}{\Talpha_2 + \Ttheta_2}$\\ \hline
    \end{tabular}
    \caption{ Limiting value of period estimate for MFL and IT-MFL models. }
    \label{tab:limEstimates}
\end{table*}

Comparing the MFL and IT-MFL period estimates, it is immediate that $t_1$ is always larger in the IT-MFL model and that its contribution to the period vanishes as $\rho_1 \to \infty$. Hence in the large $\rho_1$ limit, the period is approximately equal to $t_2$. The following two remarks give some interpretation to the extra terms that appear in the expression for $t_2$ in the case of the IT-MFL model.
\begin{remark}\label{rem:first}
$t_2$ depends linearly on $\rho_1$ with a slope given by
\begin{align}
    \frac{\partial t_2^{\textrm{MFL}}}{\partial \rho_1} &=  \frac{1}{\Ttheta_1(1-\rho_0)},\label{eqn:slopeDerMFL}\\
    \frac{\partial t_2^{\textrm{IT-MFL}}}{\partial \rho_1} &= \frac{1}{\Ttheta_1\left( 1- \rho_0 + \frac{\Talpha_2}{\Ttheta_2} (1-\rho_2) \right)\left( 1 + \frac{\Talpha_2}{\Ttheta_2}\right)}\label{eqn:slopeDerITMFL}.
\end{align}
In the case that $\rho_0<1$ and $\rho_2 <1$, both $t_2^{\textrm{MFL}}$ and $t_2^{\textrm{IT-MFL}}$ are monotonically increasing in $\rho_1$ with
\begin{equation}
    \frac{\partial t_2^{\textrm{IT-MFL}}}{\partial \rho_1} \leq \frac{\partial t_2^{\textrm{MFL}}}{\partial \rho_1}.
\end{equation}   
\end{remark}
An expression of the form $1-\rho_0>0$ may be rewritten in dimensionful parameters as $\delta_r \rho_A > \beta \rho_f$. So we see the sign and magnitude of our approximation for the slope are determined by how the timescales of activator and target production compare to one another.
Also note that the right hand side of Eq.~\eqref{eqn:slopeDerITMFL} becomes larger or smaller relative to the corresponding expression in the MFL model in Eq.~\eqref{eqn:slopeDerMFL} depending on if $\rho_2<1$ or $\rho_2>1$.

In the case where $\rho_0<1$ and $\rho_2<1$, we obtain a stronger result where the period of the IT-MFL model is controlled by the period of the MFL model up to a constant.
\begin{remark}\label{rem:second}
If $\rho_0<1,$ $\rho_2 <1,$ and $\rho_1>\max(\rho_0,\rho_2)$ then
$$t_2^{\textrm{IT-MFL}} \leq t_2^{\textrm{MFL}} + C$$
with $C = \frac{\rho_1 - \rho_2}{\Ttheta_1} \min\left( \frac{\Talpha_2}{\Ttheta_2(1-\rho_0)} ,\frac{1}{1-\rho_2}\right)$.
\end{remark}
A proof of Remark~\ref{rem:second} is given in the supplemental material. One can interpret these results as follows: in the current approximation, $t_2^{\textrm{IT-MFL}}< t_2^{\textrm{MFL}}$ when $\rho_1$ is large enough by Remark~\ref{rem:first}, and although we may find $t_2^{\textrm{MFL}}< t_2^{\textrm{IT-MFL}}$ for moderate $\rho_1$, this is controlled by the constant $C$ given in Remark~\ref{rem:second}. 
\begin{figure}
    \centering
    \includegraphics{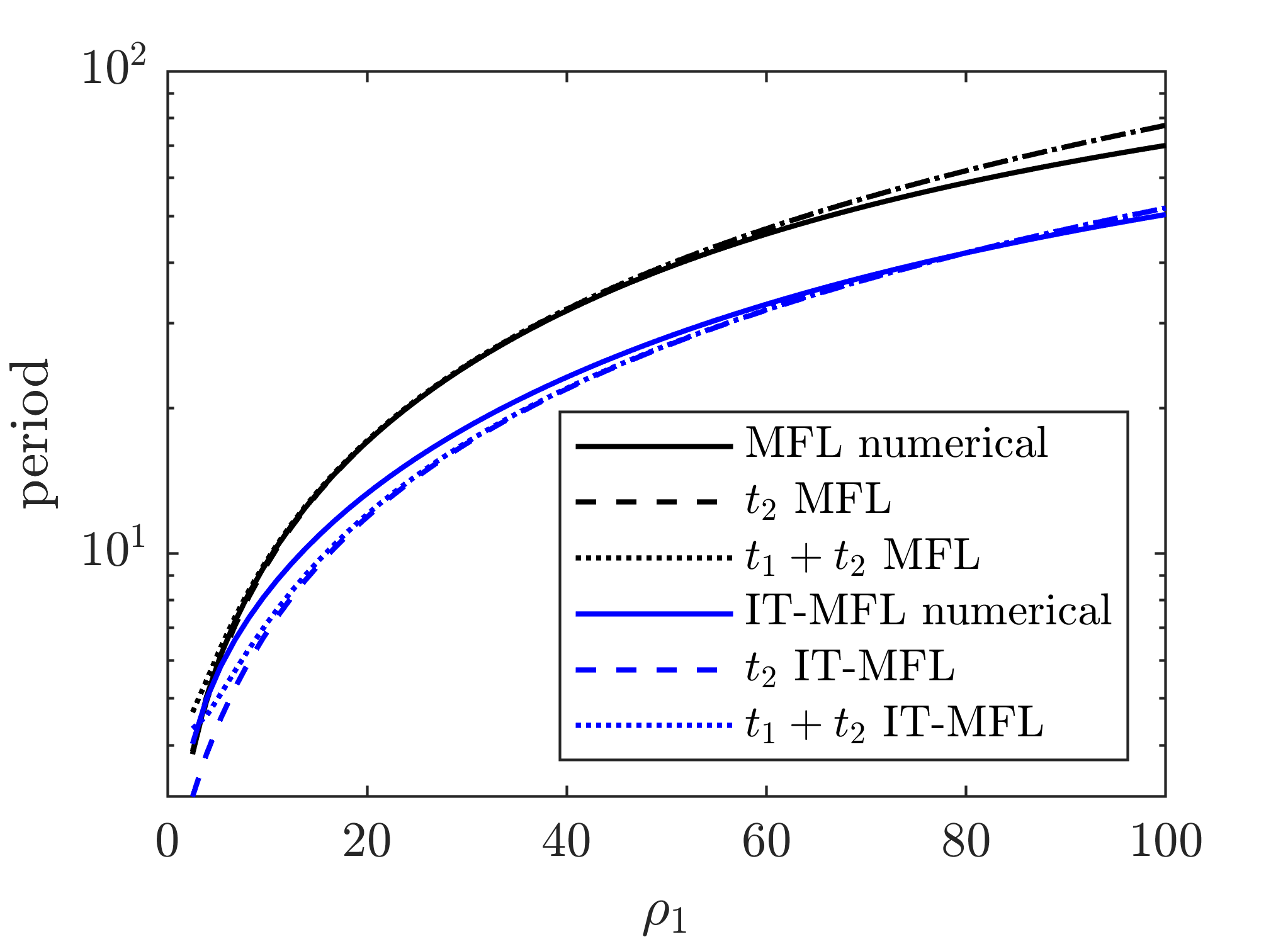}
    \caption{(Colour online). Comparison of numerically estimated period and the limiting values given in Table \ref{tab:limEstimates}. We see  $t_2^{\textrm{IT-MFL}}< t_2^{\textrm{MFL}}$ when $\rho_1$ is large, as proved in Remark \ref{rem:first}. Simulation parameters: $\delta=10^{-4},$ $\rho_0=0$,  $\Ttheta_1=1.33$, $d_a=0.3$, $d_b=0$, $\mu =0.31$, $A_0 = 4$, $\rho_2 = 0.5$, $\Talpha_2=1$, $\Ttheta_2 =1$.}
    \label{fig:limitingper}
\end{figure}
\subsection{Relation between the Kim-Forger, Goodwin, MFL and IT-MFL models}\label{sec:relation}
So far our analysis has been an extension of the original MFL paper of Fran\c{c}ois and Hakim. We now shift our attention to the work of Kim and Forger. The Kim-Forger model is a three species negative feedback loop with nonlinearity introduced by means of a sequestration function \cite{buchler2008molecular}
\begin{align}
   f(P;A,K_d)        &= \frac{1}{2} \Bigl(1 - P/A - K_d/A \notag\\
   &\quad +  \sqrt{ \left( 1 - P/A - K_d/A \right)^2 + 4K_d/A }\Bigr)  \label{eqn:seqMFL},
\end{align}
with governing equations of the form
\begin{align}
   \frac{dM}{dt}     &= \alpha_1 f(P) - \beta_1 M \label{eqn:firstOfKFPaper}, \\ 
   \frac{dP_\textrm{c}}{dt} &= \alpha_2 M - \beta_2 P_\textrm{c} ,\\ 
   \frac{dP}{dt}     &= \alpha_3 P_\textrm{c} - \beta_3 P \label{eqn:lastOfKFPaper}.
\end{align}
Kim and Forger interpret $M, P_\textrm{c},$ and $P$ as the mRNA, cytosolic, and nuclear concentrations of the target protein PER. The parameter $A$ in Eq.~\eqref{eqn:seqMFL} represents the total concentration of activator CLOCK-BMAL1 and $K_d$ is the dissociation constant of the activator-target binding reaction. Our aim in this section is to explain how one can start with the MFL model and arrive at the Kim-Forger model. We also discuss how the standard assumptions can be relaxed to provide a better approximation of the IT-MFL model. 

In the context of the MFL model Eqs.~\eqref{eqn:firstOfKFPaper}-\eqref{eqn:lastOfKFPaper} can be obtained as a consequence of the following assumptions: the transcription rate corresponding to the inactive promoter state is zero, the binding of $A$ and $B$ has reached equilibrium (rapid equilibrium approximation), the total amount of activator $A_{\tot}$ is constant, and the transcription rate of $P$ is proportional to the fraction of unbound activator $f=\frac{A}{A_{\tot}}$. By adding nuclear export and import of the target protein to the MFL model and applying the assumptions listed above, one may reduce the MFL model to Eqs.~\eqref{eqn:firstOfKFPaper}-\eqref{eqn:lastOfKFPaper}. It should be noted that the promoter states in the MFL model satisfy a conservation equation and so the proportionality assumption only holds true when the occupancy of the promoter states is constant in time. Since we are interested in the dynamics of the promoter states, we relax the assumption of constant promoter states to a quasi-steady state (QSS) approximation. The QSS approximation together with the rapid-equilibrium approximation interaction of $A$ with $B$ produce a cubic equation $\Phi_\dim([A]_\qss)=0$ for the QSS concentration $[A]_\qss$, where $\Phi_\dim$ is given by
\begin{align}
\Phi_{\textrm{dim}}([A]) = a_\dim [A]^3 + b_\dim [A]^2 + c_\dim [A] + d_\dim \label{eqn:Phidimful}
\end{align}
with coefficients
\begin{align}
a_\dim &= K_{1},\\
b_\dim &= 1+ K_{2}+ \left( g_{\tot}+B_{\tot}-A_{\tot}+K_{d}\right)K_{1}, \\
c_\dim &=(1+K_{2})(B_{\tot}-A_{\tot}+K_{d})\notag\\
&\qquad\quad+(g_{\tot}-A_{\tot})K_{d}K_{1},\\
d_\dim &=-(1+K_{2})A_{\tot}K_{d}.
\end{align}
See the supplemental material for a derivation and analysis of $\Phi_\dim([A])$. We emphasize that this polynomial is dimensionful because its dimensionless form  which we denote simply by $\Phi(\tilde{A})$ appears in our analysis contained in the supplemental material. We denote the solution to $\Phi_{\dim}([A])=0$ by $[A]_\qss$ and obtain a reduced form of the IT-MFL model 
\begin{align}
   \frac{d[r_b]}{dt}     &=  \rho_b [g:A]_{\qss} + \rho_M [g:M]_{\qss} - \delta_r [r_b], \label{eqn:firstofKF}\\
   \frac{d[B_c]}{dt}     &= \beta_1 [r_b] - \lambda_c [B_c], \label{eqn:BcKF}\\
   \frac{d[B_{\tot}]}{dt}&=   \beta_2 [B_c] - \delta_B [B_{\tot}], \label{eqn:lastofKF}
\end{align}
where $[g:A]_{\qss}$ and $[g:M]_{\qss}$ can be expressed in terms of $[A]_{\qss}$ 
\begin{align}
    [g:A]_{\qss} &= \frac{[A]_\qss K_{1} g_{\tot}}{1 + K_2 + [A]_\qss K_{1}} ,\\
    [g:M]_{\qss} &= \frac{K_2(g_\tot - [g:A]_{\qss})}{1+K_2}\label{eqn:gMqss}.
\end{align}
Of all the assumptions involved in the derivation of Eqs.~\eqref{eqn:firstofKF}-\eqref{eqn:gMqss}, the assumption of $[A]_\tot$ being constant in time seems to be the most difficult to verify. Supp. Fig.~1 shows that the IT-MFL model is well-approximated by Eqs.~\eqref{eqn:firstofKF}-\eqref{eqn:gMqss} provided that $[A]_\tot$ is constant in time and the other assumptions used in the model reduction hold true. Following a non-dimensionalization scheme similar to that of Kim and Forger, we reduce Eqs.~\eqref{eqn:firstofKF}-\eqref{eqn:lastofKF} to
\begin{align}
\frac{d\tilde{r}     }{d\tau} &=\Tbt - \tilde{r} \label{eqn:rITMFLstart},\\
\frac{d\tilde{B}_c   }{d\tau}&= \tilde{r}-\tilde{B}_c,\\
\frac{d\tilde{B}_\tot}{d\tau}   &=\tilde{B}_c -\tilde{B}_\tot,\\
\Tbt &=\left(  \frac{\Aq \tilde{K}_{1}}{1+K_2 + \Aq \tilde{K}_{1}}\right)\left(1 - \frac{\rho K_2}{1+K_2}\right) \notag \\
&\quad\quad +\frac{\rho K_2}{1+K_2}  .\label{eqn:rITMFLend}
\end{align}
where $\tau = t \delta_r$ and $[A]_\qss=B^*\tilde{A}_\qss(\tilde{B}_\tot)$. We show in the supplemental material that $\tilde{A}_{\qss}(\tilde{B}_\tot)$ is a root of the polynomial $\Phi(\tilde{A})$ mentioned earlier in this section. The dimensionless concentrations $\tilde{r}, \tilde{B}_c$, $\tilde{B}$ satisfy 
\begin{align}
[r_b] =r_b^* \tilde{r},\quad [B_c] =B_c^* \tilde{B}_c,\quad B_{\tot} = B^* \tilde{B}_\tot
\end{align}
with scaling factors
\begin{align}
r_b^* = \frac{\rho_b g_{\tot}}{\delta_r} , B_c^* =  \frac{\beta_1 \rho_b g_{\tot}}{\delta_r^2},  B^*=\frac{\beta_2 \beta_1 \rho_b g_{\tot}}{\delta_r^3},
\end{align}
and the dimensionless parameters satisfy
\begin{align}
 \rho&=\frac{\rho_M}{\rho_b}, K_2 = \frac{\alpha_2}{\theta_2}, \tilde{K}_{1}= B^*K_{1}=B^* \frac{\alpha_1}{\theta_1},\\
 \TKd&=    \frac{K_d}{B^*}, A_\tot = B^* \tilde{A}_\tot, g_\tot = B^* \tilde{g}_\tot.
\end{align}
To reduce the number of parameters in the dimensionless model, we have also assumed all degradation rates are equal $(\delta_r = \lambda_c = \delta_B)$. In a sense that is made precise in the work of Forger \cite{forger2011signal}, the assumption of equal degradation rates maximizes the likelihood of periodic behaviour in the system. Since our analysis in the next section is concerned with the situation where oscillations cease to exist, we believe this to be a reasonable simplification to our analysis. The reduced dimensionless form of the IT-MFL model contains three dynamic variables $\tilde{r}$, $\tilde{B}_c$, and $\tilde{B}_\tot$ and six parameters $\tilde{A}_\tot, \tilde{g}_\tot, \TKd,\rho,\tilde{K}_{1},$ and $K_2$. 
\begin{figure}
    \centering
    \includegraphics{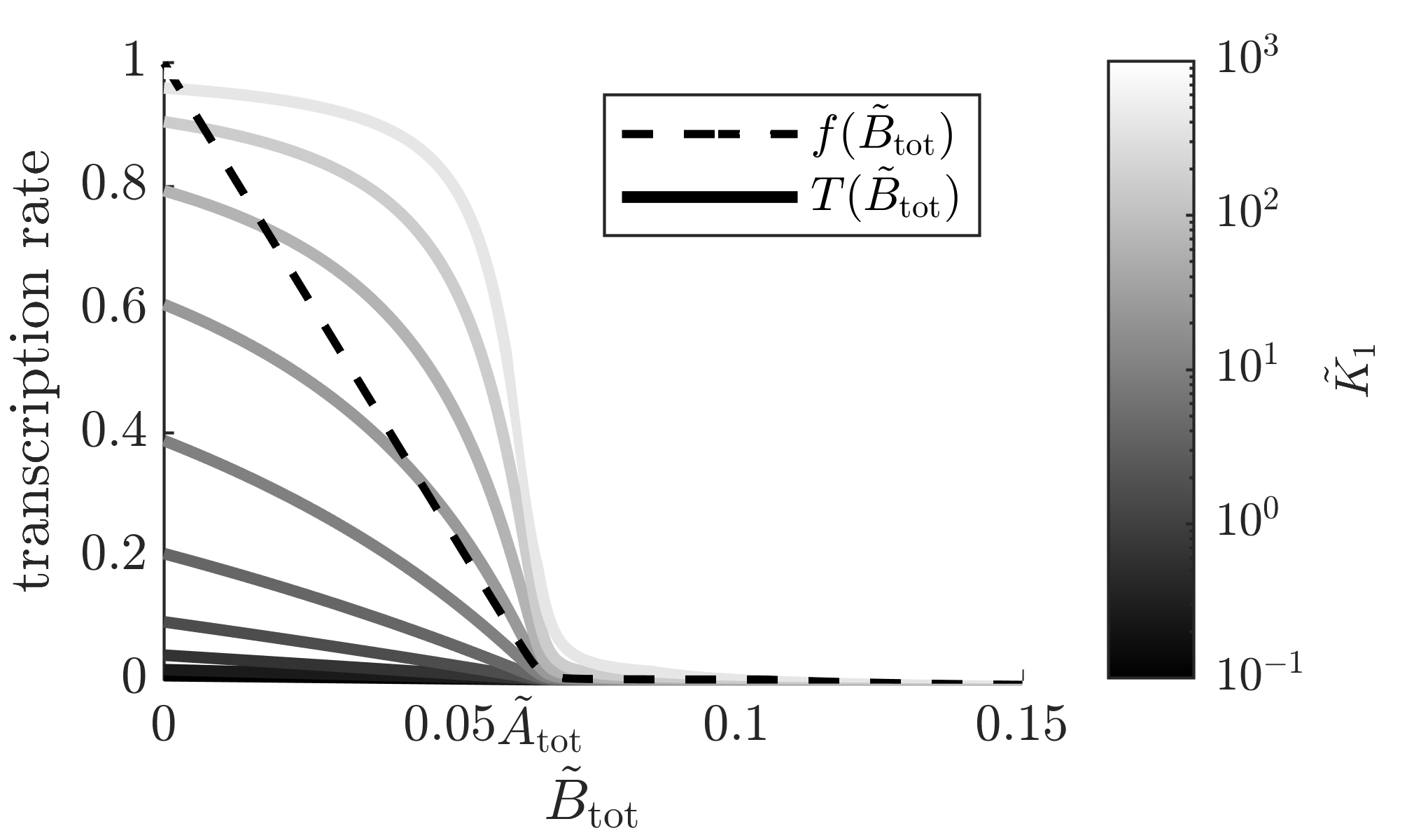}
    \caption{Comparison of the transcription functions in the dimensionless Kim-Forger and reduced IT-MFL models. The analysis of Kim and Forger shows that $f(\Bt)$ has a knee at the value $\Bt=\tilde{A}_{\tot}$, indicated on the $\Bt$-axis. As the equilibrium constant $\tilde{K}_1$ is varied, the transcription function $T(\tilde{B}_\tot)$ of the reduced IT-MFL model becomes increasingly nonlinear. The methylation parameters are null, so the increased nonlinearity should be attributed to the QSS approximation on the promoter states.  Parameters: $\tilde{A}_\tot=6.59\cdot10^{-2}$, $\tilde{K}_d=10^{-5}$, $\tilde{g}_\tot = 6.59\cdot 10^{-3}$, $\rho=K_2=0$.}
    \label{fig:Tfcompare}
\end{figure}
\subsection{Stability and period sensitivity in the reduced model}
\begin{figure}
    \centering
    \hspace*{-0.9cm}\includegraphics{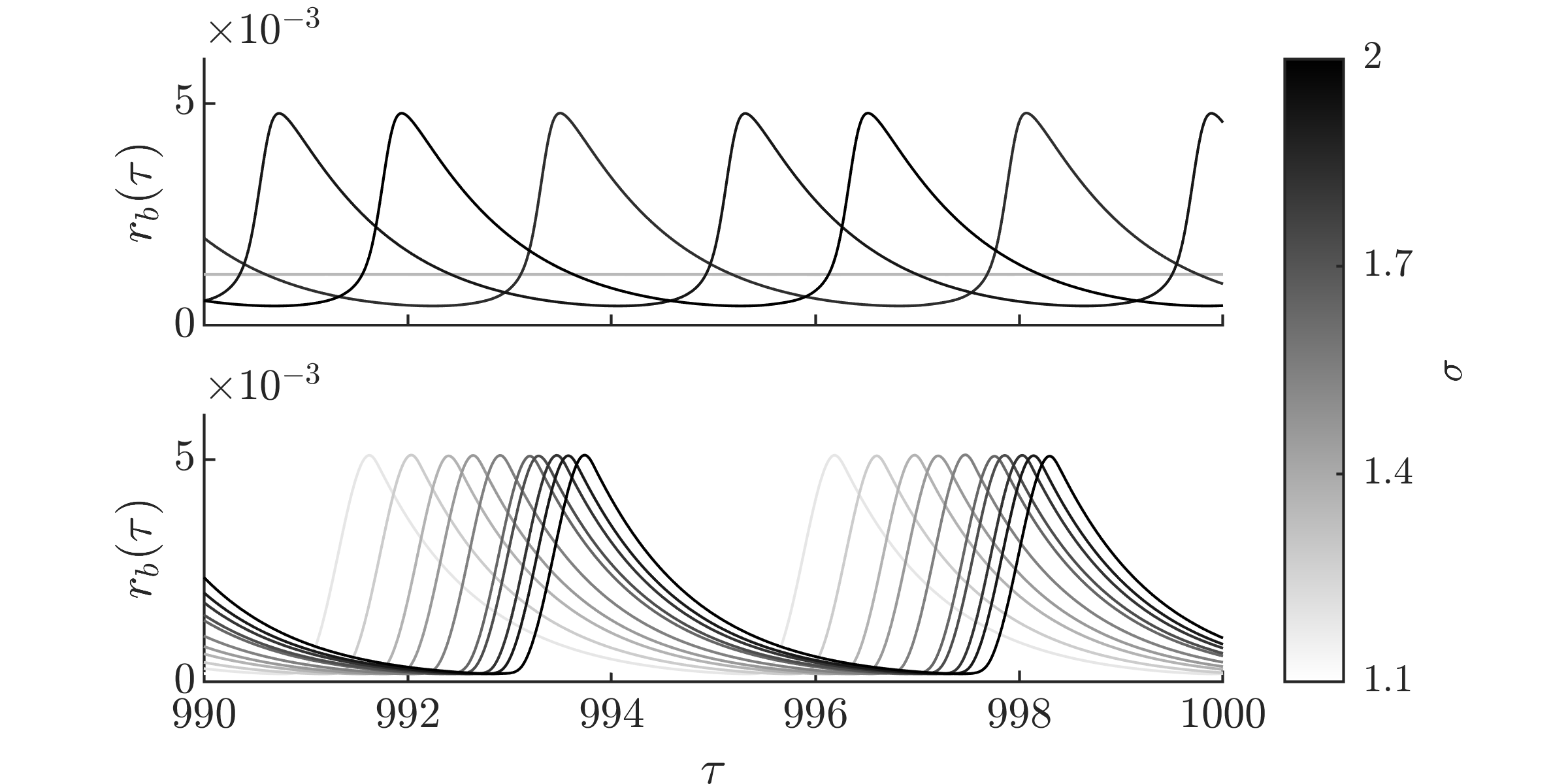}
    \caption{Simulations of the Kim-Forger and reduced IT-MFL models with initial conditions of the form $\sigma B_{\tot,\eq}$ where $B_{\tot,\eq}$ is the equilibrium value of $B_{\tot}$ in the appropriate model, and $\sigma$ varies uniformly from $\sigma=1.1$ to $\sigma=2$. (Top) The Kim-Forger model is bistable for this parameter set, so after an initial transient, solutions are either periodic or constant depending on the initial conditions. (Bottom) All solutions of the reduced IT-MFL model display periodic behaviour, regardless of their initial conditions. Parameters:  $\tilde{A}_\tot=10^{-3},\TKd=10^{-6.73},\tilde{K}_{1}=10^{2},\tilde{g}_\tot=10^{-4}$, $K_2=\rho=0$.}
    \label{fig:bistable}
\end{figure}

Written in their dimensionless forms, the reduced IT-MFL and Kim-Forger models only differ in their transcription function. Fig.~\ref{fig:Tfcompare} shows that even in the absence of methylation effects, there are substantial differences in the transcription functions of these two models. Importantly, the monotonicity of $f(\Bt)$ which is crucial to the analysis of Kim and Forger appears to be preserved when one switches from $f(\Bt)$ to $T(\Bt)$ in Fig.~\ref{fig:Tfcompare}. Under some conditions given in the supplemental material, we prove that this is indeed the case. It follows from the monotonicity of $T(\Bt)$ that there is a unique non-negative equilibrium solution to Eqs.~\eqref{eqn:rITMFLstart}-\eqref{eqn:rITMFLend}. Moreover, we show in the supplemental material that the monotonicity conditions also ensure non-negative solutions to Eqs.~\eqref{eqn:rITMFLstart}-\eqref{eqn:rITMFLend} are bounded. This implies that the reduced IT-MFL model constitutes a bounded monotone cyclic feedback (MCF) system, and so any solution must converge to static equilibrium, a non-constant periodic solution, or a combination of homoclinic and heteroclinic orbits connecting the previous two types of equilibria. This follows from the generalization of the Poincar\'{e}-Bendixson theorem to MCF systems by Mallet-Paret and Smith~\cite{mallet1990poincare}. Although multistability is possible for MCF systems, we see in Fig.~\ref{fig:bistable} that unlike in the Kim-Forger model, this does not appear to be the case in the reduced IT-MFL model. We verify this observation through numerical bifurcation analysis.

\begin{figure}
    \centering
    \hspace*{-0.4cm}\includegraphics{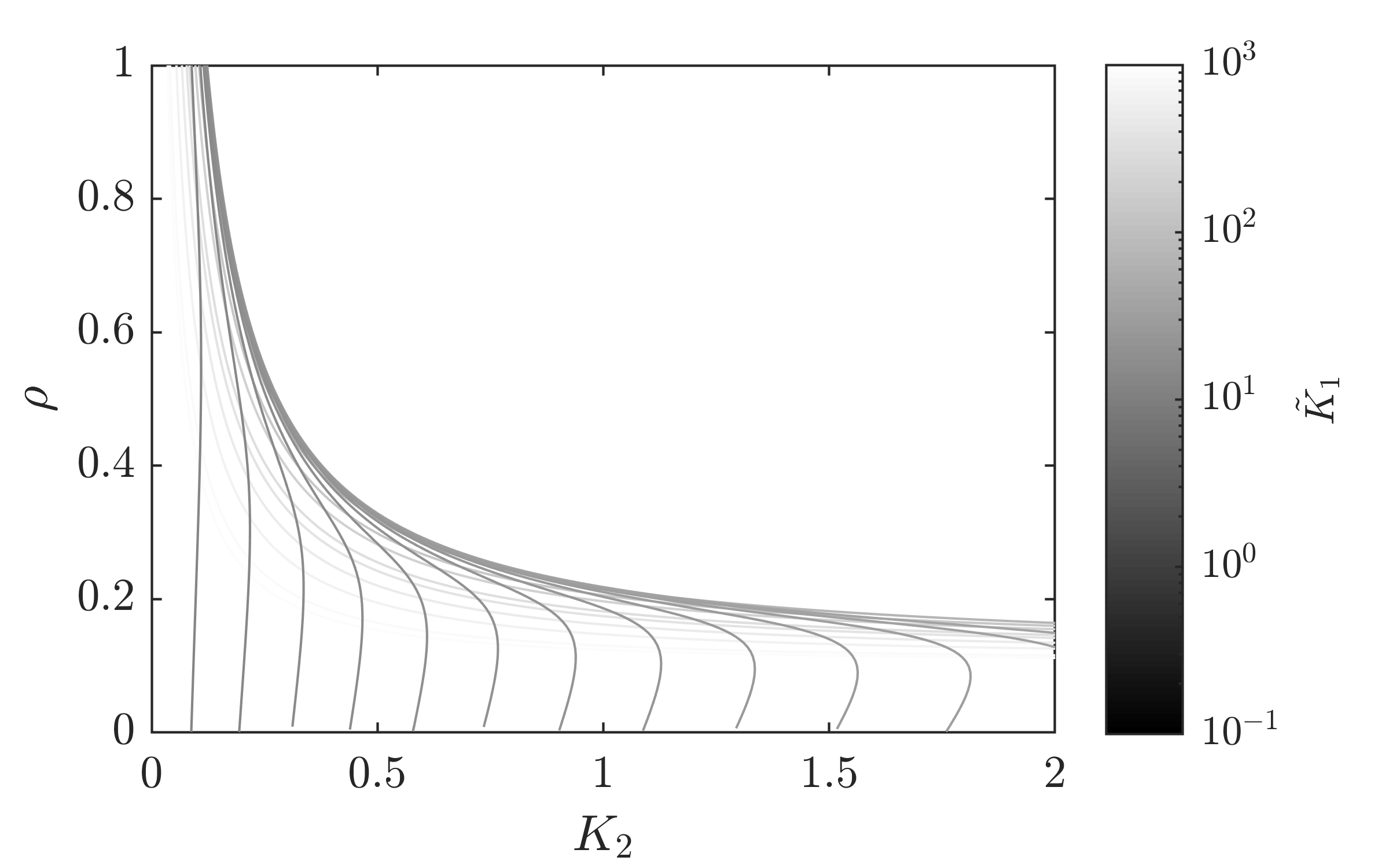}
    \caption{Each grayscale coloured curve corresponds to a Hopf bifurcation in $\rho$, $K_2$. The colour gives the value of $\tilde{K}_1$. Using the Lyapunov coefficient, we found these bifurcations are supercritical. To the left of each curve there is a stable periodic solution and to the right there is a single stable equilibrium. The other parameters are $\tilde{A}_\tot=1.31\cdot10^{-1}$, $\tilde{g}_\tot=1.31\cdot10^{-2}$, $\tilde{K}_d=10^{-5}$. This shows that methylation can make the clock arrhythmic. Hopf bifurcations do not occur for $K_1$ below approximately $11.6$ as the bifurcation value of $K_2$ goes to zero.}
    \label{fig:bifur}
\end{figure}

We fix parameters $\tilde{A}_\tot$,$\TKd$, and $\tilde{g}_\tot$, and numerically compute the location of any Hopf bifurcations in the methylation parameters $\rho$ and $K_2$. We use pseudo-arclength continuation~\cite{allgowergeorg2003continuation} of the equations $\Phi(\tilde A_\qss)=0$, $T(B_{\tot})=B_{\tot}$ (condition for equilibrium), and $-\frac12\sqrt[3]{T'(B_{\tot})}-1=0$ (complex conjugate pair of eigenvalues of the equilibrium's linearization are crossing the imaginary axis) in the unknowns $\tilde A_\qss$, $B_{\tot}$, $\rho$, and $K_2$. The value of $\tilde{K}_{1}$ is varied from $10^{-1}$ to $10^{3}$ and the result is shown in Fig.~\ref{fig:bifur}. These Hopf bifurcations are confirmed to be supercritical by numerically evaluating the first Lyapunov coefficient~\cite{kuznetsov2013bifurcation}. Therefore, to the left of each curve, there is a stable periodic solution and an unstable equilibrium. To the right of each curve, there is a single stable equilibrium. These findings are in agreement with Fig.~\ref{fig:bistable}, where bistability appears to be absent from the reduced IT-MFL model. Notice the Hopf bifurcation curve disappears for $\tilde{K}_{1}$ below a specific value (11.6 for the parameters in Fig.~\ref{fig:bifur}) as $K_2 \to 0$. Also, the reduced IT-MFL exhibits oscillations for sufficiently high $\tilde{K}_{1}$ and for sufficiently low methylation parameters $\rho$ and $K_2$. The influence of the methylation parameters on the location of the Hopf bifurcation can also be observed when one performs parametric sensitivity analysis on the period.

\begin{figure}
    \centering
    \hspace*{-0.5cm}\includegraphics{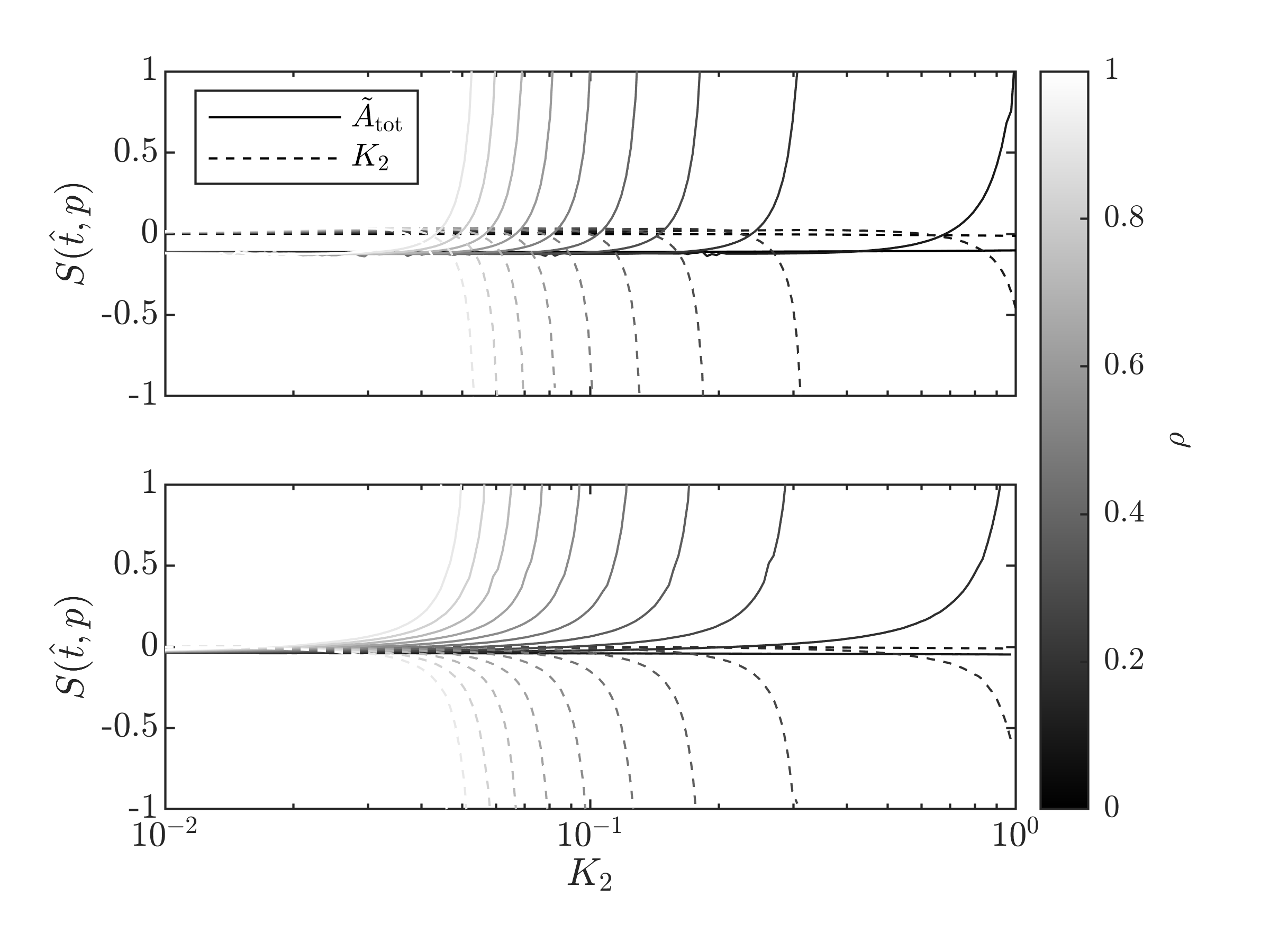}
    \caption{Numerically computed period sensitivities. We use $S(\hat{t},p):=\frac{p}{\hat{t}}\frac{\partial \hat{t}}{\partial{p}}$ to denote the sensitivity of the period $\hat{t}$ with respect to a parameter $p$. The parameter $\rho$ was varied uniformly from $0$ to $1$ and this is represented by the transparency of each sensitivity curve. Curves in the top panel were computed with $\tilde{g}_\tot = 10^{-3}$ and a value of $\tilde{g}_\tot=10^{-2}$ was used in the bottom panel. In general, the sensitivities of the period grow as $K_2$ is increased and the fixed point becomes stable, rendering the model arrhythmic. Simulation parameters: $\tilde{A}_\tot=6.59\cdot10^{-2},\TKd=10^{-6},\tilde{K}_{1}=10^{3}.$}
    \label{fig:persens}
\end{figure}

We use a method of sensitivity analysis intended for oscillating systems \cite{ingalls2004autonomously} which avoids some of the numerical difficulties encountered when studying period sensitivity. The sensitivities of the period with respect to $\tilde{A}_\tot$ and $K_2$ are shown in Fig.~\ref{fig:persens}. We see that variations in the methylation parameters $\rho$ and $K_2$ affect the sensitivity of the period with respect to other parameters in the model and these changes are most dramatic as the sensitivities diverge in the neighborhood of the Hopf bifurcation. Since it would be risky for a biological clock to operate close to a Hopf bifurcation and the sensitivities are relatively unaffected by the methylation parameters away from the Hopf bifurcation, we see that the period is robust to changes in the methylation parameters. As in Fig.~\ref{fig:bifur}, where higher $\rho$ values cause the bifurcation to occur for lower values of $K_2$, Fig.~\ref{fig:persens} shows that higher values of $\rho$ cause the sensitivity curves to diverge for lower $K_2$. One could interpret this finding as saying that lower $\rho$ values allow the model to tolerate more methylation (larger $K_2$ value) before becoming arrhythmic. 
\section{Discussion}
We have introduced and analyzed a mathematical model of a mixed feedback loop with an intermediate promoter state. By extending the perturbative analysis of Fran\c{c}ois and Hakim, we found that the uniqueness and stability of its equilibrium solution were preserved provided that the transcription rate of the new promoter state was between the inactive and active transcription rates. Under some additional restrictions on the parameters, we derived leading order estimates for the period in the IT-MFL model as well as bounds for the difference in period between the MFL and IT-MFL models. Following our perturbative analysis, we used a different set of assumptions on the parameters to show how a modified Goodwin oscillator model -- previously studied by Kim and Forger -- can be obtained as an approximation of the IT-MFL model. Working in a generalization of this approximation, we found that although methylation influences the period and its sensitivity to other parameters in the reduced IT-MFL model, excessive methylation can remove oscillations from the system. Since the assumptions for the reduced IT-MFL model are more general than those of the Kim-Forger model, we believe a systematic comparison of these two models viewed as approximations of the full IT-MFL model would be a valuable contribution to the literature. 

Throughout our perturbative analysis, the model's qualitative behaviour was dictated by the promoter state with the highest transcription rate. In future work, one could test if this finding generalizes to an $n$-promoter state mixed feedback loop model. In addition, incorporating DNA methylation machineries, namely DNA methyltransferases (DNMTs) and Ten-eleven translocation enzymes (TETs), as well as cellular states that influence DNA modifications (e.g., development, aging, or even cancer) may provide a better representation of biological reality. These features may introduce sufficient nonlinearity for the model to stay rhythmic at higher methylation levels, allowing methylation to play a greater role in controlling the periodic behaviour \cite{greenberg2019diverse}. Although such a model would likely be too detailed for the type of analysis we have used here, we hope our results will provide a useful starting point for analyzing such an extension. Future work will attempt to explore the role of oscillatory patterns in DNA modifications seen in cell fate determination, aging, and disease \cite{OhPetronis2021,parry2020active}.

Molecular noise is another biologically important factor \cite{raj_nature_2008,stinchcombe_population_2012} which could be studied in future work. Recent work of Karapetyan and Buchler on a stochastic generalization of the MFL model \cite{karapetyan2015role} and the work of Wang and Peskin on the effects of molecular noise on entrainment in an MFL model of the circadian clock \cite{wang_entrainment_2018} provide useful starting points for extending our analysis to the stochastic setting. More generally, several stochastic models of methylation \cite{utsey2020mathematical,ancona2020competition,busto-moner_stochastic_2020} and histone \cite{sood2020quantifying,sood2021quantifying,bhattacharyya2020stochastic} dynamics, as well as biophysical epigenetic models \cite{jost2014bifurcation,cortini2016physics} have appeared in recent years. As it becomes more common practice to include epigenetic effects in gene regulatory network models -- for example \cite{Chen2019.12.19.883280,jia2019possible,flottmann2012stochastic,folguera2019multiscale} -- we anticipate the detailed standalone models of epigenetic dynamics will come to be useful in gene regulatory network models. It has been argued that the incorporation of epigenetic factors in mathematical models of gene regulatory networks is one of the most important challenges in the development of large-scale predictive models of post-embryonic systems \cite{rothenberg_causal_2019}. As these models come into existence, we expect studies of reduced models will continue to provide unique insights into the intricate machinery of regulated gene expression, unavailable through detailed model simulation alone.

\section{Acknowledgments}
ARS acknowledges the support of the Natural Sciences and Engineering Research Council of Canada (NSERC): RGPIN-2019-06946.

\bibliography{references.bib}
\end{document}


\maketitle
\section{Proofs of propositions from the main document}\label{sec:proofs}
\setcounter{remark}{1}
\begin{prop}\label{prop:only} If $\rho_0 < \rho_2 \leq \rho_1$ then $\lambda_5^{\textrm{IT-MFL}} <0$ to dominant order as $\delta \to 0$.
\end{prop}
\begin{proof}
Let $\lambda \in \bR$ such that $\lambda_5^{\textrm{IT-MFL}} = \lambda + \mathcal{O}(\delta)$ as $\delta \to 0$. From Table IV of the main document, we obtain 
\begin{align}\label{eqn:firstofproof}
    \lambda  &= \frac{ \left(1 + \kmeth \right)\rho_1-\left( \rho_0 + \kmeth\rho_2\right) }{\frac{1}{\Ttheta_2}\left( \rho_{0}-\rho_{1}\right) }
    = \frac{- \left(\rho_0 - \rho_1 \right) + \frac{\Talpha_2}{\Ttheta_2}\left(\rho_1 - \rho_2 \right)  }{ \frac{1}{\Ttheta_2} \left( \rho_0 - \rho_1\right)}  = -\frac{1}{\Ttheta_2} + \frac{\Talpha_2}{\Ttheta_2^2} \left(\frac{\rho_1-\rho_2}{\rho_0-\rho_1} \right).
\end{align}
we see $\lambda<0$ if and only if
\begin{equation}\label{eqn:stabineq}
  \frac{\Talpha_2}{\Ttheta_2} \left( \frac{\rho_1 - \rho_2}{\rho_0 - \rho_1} \right) < 1.
\end{equation}
It is always the case that $\frac{\Talpha_2}{\Ttheta_2}> 0$ and $\frac{\rho_1 - \rho_2}{\rho_0 -\rho_1}\leq0$ follows from the assumption that $\rho_0 < \rho_2 \leq \rho_1$. So as long as $\rho_0 < \rho_2 \leq \rho_1$, the left hand side of Eq.~\eqref{eqn:stabineq} is non-positive and the inequality holds true.
\end{proof}
\begin{remark}\label{rem:second}
If $\rho_0<1,$ $\rho_2 <1,$ and $\rho_1>\max(\rho_0,\rho_2)$ then
$t_2^{\textrm{IT-MFL}} \leq t_2^{\textrm{MFL}} + C$
with $C = \frac{\rho_1 - \rho_2}{\Ttheta_1} \min\left( \frac{\Talpha_2}{\Ttheta_2(1-\rho_0)} ,\frac{1}{1-\rho_2}\right)$.
\end{remark}
\begin{proof}
First, notice that $\left(1 + \frac{\Talpha_2}{\Ttheta_2} \right)\geq 1$, and since we have assumed $\rho_0 <\rho_1$ and $\rho_2<\rho_1$
\begin{align}
    t_2^{\textrm{IT-MFL}} &=  2 +  \frac{\rho_1-\rho_0 + \frac{\Talpha_2}{\Ttheta_2}\left(\rho_1- \rho_2 \right)}{\Ttheta_1\left( 1 - \rho_0 + (1-\rho_2)\frac{\Talpha_2}{\Ttheta_2}\right)\left(1 + \frac{\Talpha_2}{\Ttheta_2} \right) } 
    \leq 2+  \frac{\rho_1-\rho_0 + \frac{\Talpha_2}{\Ttheta_2}\left(\rho_1- \rho_2 \right)}{\Ttheta_1\left( 1 - \rho_0 + (1-\rho_2)\frac{\Talpha_2}{\Ttheta_2}\right) }\label{eqn:firstineq}.
\end{align}
Using $ 1-\rho_0 + (1-\rho_2)\frac{\Talpha_2}{\Ttheta_2} \geq  1-\rho_0$ and Eq.~\eqref{eqn:firstineq}, we find that 
\begin{align}
t_2^{\textrm{IT-MFL}}    &\leq 2 + \frac{\rho_1 - \rho_0}{\Ttheta_1(1-\rho_0)} + \frac{1}{ \frac{\Ttheta_1\left( 1 - \rho_0 \right)}{\frac{\Talpha_2}{\Ttheta_2}\left(\rho_1- \rho_2 \right)} + \Ttheta_1 \left( \frac{1-\rho_2}{\rho_1-\rho_2}\right) } \label{eqn:secondineq}.
\end{align}
Since we assumed $\rho_0<1$ and $\rho_2 < 1$, we know $(a+b)^{-1} \leq \min(a^{-1},b^{-1})$ with $a=  \frac{\Ttheta_1\left( 1 - \rho_0 \right)}{\frac{\Talpha_2}{\Ttheta_2}\left(\rho_1- \rho_2 \right)}$ and $b=\Ttheta_1 \left( \frac{1-\rho_2}{\rho_1-\rho_2}\right)$ and so we obtain from Eq.~\eqref{eqn:secondineq} $t_2^{\textrm{IT-MFL}} \leq t_2^{\textrm{MFL}} + C$
with $C = \frac{\rho_1 - \rho_2}{\Ttheta_1} \min\left( \frac{\Talpha_2}{\Ttheta_2(1-\rho_0)} ,\frac{1}{1-\rho_2}\right)$.
\end{proof}

\section{Asymptotic analysis}\label{sec:asymptoticAnalysis}
We now provide a more thorough discussion of the perturbative calculations from Table V of the main document. We use some assumptions from the original MFL paper \cite{franccois2005core} in our derivation and only verify the validity of these assumptions by comparing the results of the analysis to a fully numerical approach.
\subsection{Limiting value of period estimate}
In the special case that $d_b=0$ and $\rho_1$ is large, Fran\c{c}ois and Hakim derive an approximate expression for the period of the limit cycle. We take the same approach with the IT-MFL model. After solving the governing equations for each phase of the limit cycle from the first two rows of Table V, we then approximate the value of $r_1$ by solving $r_{II}(t_2)=r_1$ and neglecting terms that are exponentially small as functions of time. This gives the $r_1$ estimate in Table VI. Applying the same strategy to the equation $g_{M,II}(t_2)=g_{M1}$ gives the $g_{M1}$ estimate in Table VI.

Next, an approximation for $t_1$ can be obtained from the equation $a_{I}(t_1) = 0$. Since $t_1$ is small when $\rho_1$ is large, we may Taylor expand $a_{I}(t_1)$ to second order in $t_1$ and obtain the approximate expression in Table VI. We obtain a value of $g_{M2}$ by expanding the equation $g_{M,I}(t_1)= g_{M2}$ to zeroth order in powers of $\rho_1^{-1}$ to obtain $g_{M2}\approx g_{M1}$. Approximating $r_I(t_1) = r_2$ to zeroth order in $\rho_1^{-1}$ gives the $r_2$ estimate from Table VI. Finally, an approximate expression for $t_2$ can be obtained by solving $b_{II}(t_2)=0$ and neglecting terms that decay exponentially as functions of $t_2$.

\subsection{Higher order corrections}
We follow the same approach as Fran\c{c}ois and Hakim to derive the $\mathcal{O}(\sqrt{\delta})$ period estimate of the period in the IT-MFL model. Two boundary layers appear at the Phase I-II boundary. The first layer denoted by $\textrm{BL}_1$ occurs prior to $t_1$ when the quasi-steady state approximation of $g$ breaks down due to the decline in $A$. We assume $\textrm{BL}_1$ begins when $g$ and $a$ are of the same magnitude so that $g \sim a \sim\frac{\delta}{a}$ and thus $a \sim g \sim \sqrt{\delta}$ within the boundary layer. This scaling suggests that $\textrm{BL}_1$ is of thickness $t_1 -t \sim \sqrt{\delta}$. We verify this by determining the dominant order contributions to the governing equation for $g$. We define the rescaled time $\tau := \frac{t}{\eta(\delta)}$ and rescaled protein concentrations $\hat{g} := \frac{g}{\sqrt{\delta}}$, $\hat{a}:= \frac{a}{\sqrt{\delta}}$, and since we are in the high-$A$-phase, $a=\delta A$. Making the appropriate substitutions, we obtain
\begin{align}
 \frac{\sqrt{\delta}}{\eta}\frac{d\hat{g}}{d\tau}  = \Ttheta_1 \left( (1-\sqrt{\delta}\hat{g}-g_M) + \frac{\Ttheta_2}{\Ttheta_1}g_M - \frac{\hat{g} \hat{a}}{A_0} \right) - \sqrt{\delta}\Talpha_2 \hat{g}.\label{eqn:fulleta}
\end{align}
If we assume $\eta(\delta) = \sqrt{\delta}$ then terms of the form $\sqrt{\delta}\hat{g}$ on the right hand side of Eq.~\eqref{eqn:fulleta} can be neglected. Transforming back to the original variables $(t,g,a)$, the dominant order terms in Eq.~\eqref{eqn:fulleta} become
\begin{align}
   \dot{g} = \Ttheta_1 \left( 1-g_M + \frac{\Ttheta_2}{\Ttheta_1}g_M -\frac{g a}{\delta A_0}  \right).\label{eqn:reducedeta}
\end{align}
Since the time-scaling factor $\eta$ is $\mathcal{O}(\sqrt{\delta})$, we see that $\textrm{BL}_1$ is indeed of thickness $\sqrt{\delta}$. To solve Eq.~\eqref{eqn:reducedeta}, we approximate $a(t)$ by its linearization at the right endpoint of the boundary layer to obtain
\begin{align}
   a(t) = (1-r_2)(t_1-t) + o(\sqrt{\delta}) \label{eqn:aapprox} 
\end{align}
More precisely, we have used the linearization of $a(t)$ with its derivative approximated to leading order in $\delta$ to obtain Eq.~\eqref{eqn:aapprox}. Substitute Eq.~\eqref{eqn:aapprox} into Eq.~\eqref{eqn:reducedeta} and integrate to obtain
\begin{align}
   g(t_1) - g(t) e^{-\kappa_2 (t_1-t)^2} = \left(\Ttheta_1 + (\Ttheta_2 - \Ttheta_1)g_{M2}\right) \int_{t-t_1}^{0} e^{\kappa_2 u^2}du,
\end{align}
in which $\kappa_2 = \frac{\Ttheta_1 (1-r_2)}{2A_0 \delta}$. Taking the $t\to - \infty$ limit gives the value of $g$ at the Phase I-II boundary $g_I(t_1) = g_{II}(0)$, given in the second last row of Table V. The analysis of the $\textrm{BL}_4$ boundary layer is similar to the case of $\textrm{BL}_1$. We refer the reader to the appendix of the original MFL paper for further details on the derivation.

\section{Analysis of the reduced model}\label{sec:reducedAnalysis}
\subsection{Derivation of reduced IT-MFL model}
\begin{figure}
    \centering
    \includegraphics{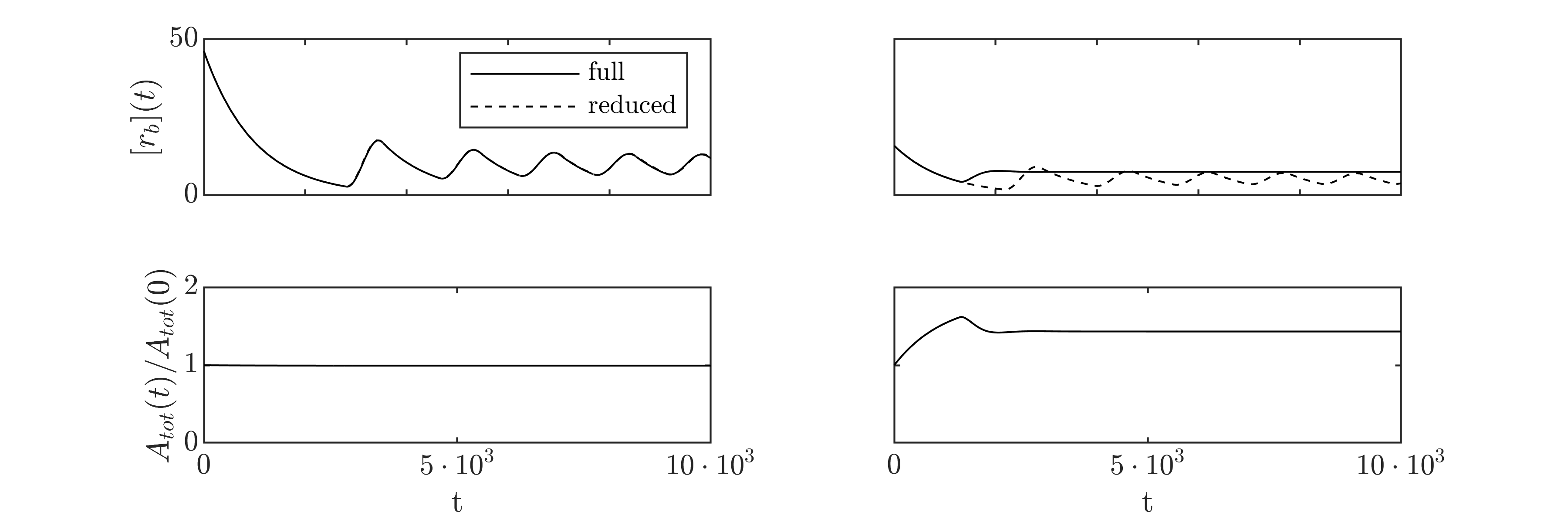}
    \caption{Comparison of reduced and full IT-MFL models. (Left column) When $A_{\tot}(t)$ is approximately constant in time and the timescale separations used in the model reduction hold true, there is strong agreement between numerical solutions to the reduced and full IT-MFL models. (Right column) By increasing the decay rate $\delta_A$ so that $A_{\tot}$ is no longer constant in time, we see the qualitative behaviour differs between the reduced and full IT-MFL models. Simulation parameters: $\alpha_1=10^{-2}$, $\theta_1=10^2$, $\alpha_2=0$, $\theta_2=1$, $\rho_M=0$, $\rho_f=0$, $\rho_b=10$, $\beta_1=\beta_2=\lambda_c=10^{-2}$, $\rho_A=10^{-1}$, $\gamma_{+}=10^2$, $\gamma_{-}=10^{-2}$, $\delta_B=\delta_{AB}=10^{-3}$. For the left column $\delta_A=10^{-3}$ and for the right column $\delta_A=10^{-2}$.}
    \label{fig:timestep_reduced_full_agree}
\end{figure}

We relax the assumption of constant promoter states to a quasi-steady state (QSS) approximation. Using the QSS approximation and the rapid equilibrium approximation for the binding of $A$ and $B$, we obtain
\begin{align}
   K_d &= \frac{[A] [B]}{[A:B]} \label{eqn:ITMFLstart}\\
   A_{\tot}&= [A] + [g:A] + [A:B]\\
   B_{\tot}&= [B] + [A:B]\\
   [g:A]&= \left( g_{\tot} - [g:A] - [g:M]\right) K_{1}[A]\\
    [g:M]&= K_2 (g_{\tot} - [g:A] - [g:M]) \label{eqn:ITMFLend}.
\end{align}
where $K_d = \frac{\gamma_{-}}{\gamma_{+}}$, $K_{1} = \frac{\alpha_1}{\theta_1}$, $K_2=\frac{\alpha_2}{\theta_2}$ and the promoter states obey the conservation $g_{\tot} = [g] + [g:A]+[g:M]$. We make the assumption that $A_\tot$ is constant in time, but no such assumption is made about $B_\tot$. Eqs.~\eqref{eqn:ITMFLstart}-\eqref{eqn:ITMFLend} are invariant under the rescaling $K_{1} \to \tilde{K}_{1}/B^*$ and $x\to B^* \tilde{x}$ for all other dimensionful parameters and dynamic variables, where $B^*=\frac{\beta_2 \beta_1 \rho_b g_{\tot}}{\delta_r^3}$. We have chosen this rescaling so that it coincides with the rescaling used by Kim and Forger in their non-dimensionalization of their model \cite{kim2012mechanism}. In the rescaled variables, Eqs.~\eqref{eqn:ITMFLstart}-\eqref{eqn:ITMFLend} reduce to a cubic equation for $\tilde{A}$
\begin{align}
0&=\Phi(\tilde{A}) = a A^3 + b A^2 + c A +d 
\end{align}
with coefficients
\begin{align}
a &= \tilde{K}_{1} \label{eqn:a}\\
b &=K_2+\tilde{K}_{1}\tilde{g}_{\tot}-\tilde{A}_{\tot}\tilde{K}_{1}+\tilde{B}_{\tot}\tilde{K}_{1}+\tilde{K}_{d}\tilde{K}_{1}+1 \\
c &=\tilde{B}_{\tot}-\tilde{A}_{\tot}+\tilde{K}_{d}-\tilde{A}_{\tot}K_2+\tilde{B}_{\tot}K_2+K_2\tilde{K}_{d}-\tilde{A}_{\tot}\tilde{K}_{d}\tilde{K}_{1}+\tilde{K}_{d}\tilde{K}_{1}\tilde{g}_{\tot}\\
d &=-\tilde{A}_{\tot}\tilde{K}_{d}-\tilde{A}_{\tot}K_2\tilde{K}_{d}\label{eqn:d}
\end{align}
Although we cannot guarantee that $\Phi(A)$ has a unique real root, we found in our numerical studies that there was always a unique non-negative real root for the parameter sets we considered. Since $\tilde{B}_{\tot}$ is the only dynamic variable appearing in the coefficients of $\Phi(\tilde{A})$, we denote the non-negative real root of $\Phi(\tilde{A})$ by $\tilde{A}_{\qss}(\tilde{B}_\tot)$. The assumption that the degradation rates $\delta_A$ and $\delta_{AB}$ are equal is our final step in reducing the IT-MFL model to a modified Goodwin oscillator. In this case, it is sufficient to only track $B_{\tot}$ instead of $[B]$ and $[A:B]$, and we obtain Eqs.~(33)-(35) of the main document. As already mentioned, the cytosolic protein state $[B_c]$ was not present in the IT-MFL model and has been added so that the reduced form of the model will be a three-stage rather than two-stage feedback loop. We see in Supp. Fig.~\ref{fig:timestep_reduced_full_agree} that there is good agreement of the full and reduced models when $A_{\tot}$ is close to constant in time, but qualitative differences appear when this is not the case.

\subsection{Monotonicity of the new transcription function}
The following conditions guarantee monotonicity of the transcription function $\Tbt$. Notice that the second condition $(\rho<1)$ is a requirement that the new promoter state must have a transcription rate intermediate to the active and inactive states.
\begin{prop}[Monotonicity of transcription function]
If $\tilde{B}_{\tot} > (1+ \frac{\TKd \tilde{K}_{1}}{K_2}) \tilde{A}_{\tot}$ and $\rho<1$ then the transcription function $\Tbt$ is monontically decreasing.
\end{prop}
\begin{proof}
Differentiate $\Phi(\tilde{A})=0$ with respect to $\tilde{B}_{\tot}$ to find
\begin{align}
0= c_1 \frac{d\tilde{A}_{\qss}}{d\tilde{B}_\tot}+ c_0\label{eqn:implicit}
\end{align}
with coefficients
\begin{align}
    c_1 &= \tilde{B}_{\tot}-\tilde{A}_{\tot}+\tilde{K}_{d}+2\Aq\left(K_{2}+\tilde{K}_{1}\tilde{g}_{\tot}-\tilde{A}_{\tot}\tilde{K}_{1}+\tilde{B}_{\tot}\tilde{K}_{1}+\tilde{K}_{d}\tilde{K}_{1}+1\right)\notag\\
        &\quad+3\tilde{K}_{1}\Aq^2 + K_2\left(B_{\tot}- A_{\tot}\left(1 + \frac{\tilde{K}_d\tilde{K}_{1}}{K_2}\right)\right)+K_{2}\tilde{K}_{d}+\tilde{K}_{d}\tilde{K}_{1}g_{\tot}\\
    c_0 &= (K_2 + 1)\Aq + \tilde{K}_{1}\Aq^2 
\end{align}
Notice that $c_0>0$ and $c_1>0$ provided that $\tilde{B}_{\tot} > (1+ \frac{\TKd \tilde{K}_{1}}{K_2}) \tilde{A}_{\tot}$. Hence we require $\tilde{A}_\qss'(\Bt) < 0$ in order for Eq.~\eqref{eqn:implicit} to hold true. Next notice that for $a_2>0$, the function
\begin{align}
    g(x;a_1,a_2,a_3) = a_1 + \frac{a_2 x}{1+a_3 x}
\end{align}
is monontically increasing and we may write 
\begin{equation}
    \Tbt = g(\Aq;a_1,a_2,a_3) \label{eqn:gbt}
\end{equation}
for some $a_1,a_2,a_3 \in\bR$. Inspection of the transcription function reveals
\begin{align}
    \Tbt =\left(  \frac{\Aq \tilde{K}_{1}}{1+K_2 + \Aq \tilde{K}_{1}}\right)\left(1 - \frac{\rho K_2}{1+K_2}\right) +\frac{\rho K_2}{1+K_2}\label{eqn:tf}
\end{align}
so $\rho<1$ implies $a_2>0$ in Eq.~\eqref{eqn:gbt}. Consequently $T'(\Bt) = \frac{dg}{d\tilde{A}_\qss} \frac{d\tilde{A}_\qss}{d\Bt} < 0$.
\end{proof}
\subsection{Boundedness of the reduced IT-MFL in the monotonic case}
The monotonicity assumption of the previous section ($\tilde{B}_{\tot} > (1+ \frac{\TKd \tilde{K}_{1}}{K_2}) \tilde{A}_{\tot}$ and $\rho<1$) is sufficient for ensuring all non-negative solutions of the reduced IT-MFL model are bounded. We prove this by constructing a bounding box in the phase space. Evaluate the normal derivatives on the $r=0$, $B_c=0$, $B=0$ planes to find
\begin{align}
    \dot{\tilde{r}} &= \Tbt \geq 0,\\
    \dot{\tilde{B}}_c   &=  \tilde{r} \geq 0 ,\\
    \dot{\tilde{B}}_\tot     &= \tilde{B}_c \geq 0,
\end{align}
respectively, and so solutions flow into the $\tilde{r}>0,\tilde{B}_c>0,\tilde{B}_\tot>0$ region. To obtain an upper bound, recall that $\Tbt$ is monotonically decreasing and so $\Tbt<T(0)$ for all $\tilde{B}_{\tot}>0$. Notice that if $\tilde{r}>T(0)$ then $\dot{\tilde{r}}<0$. This shows that a solution starting in the region $0< \tilde{r}(0) < T(0)$ cannot exit this region. By the same reasoning, solutions that start with $0 < \tilde{B}_c(0) < T(0)$ will maintain this property as time goes on. The same holds true of $\Bt$ and so we have obtained an upper bound for our system. It should be noted that we implicitly used the fact that $\tilde{A}_{\qss}\geq0$ to conclude $\Tbt\geq 0$. Since we have not proved existence of a non-negative root to $\Phi(\tilde{A})$, we should remark that our argument holds true.

\bibliography{references}